# Examining Periodic Solar Wind Density Structures Observed in the SECCHI Heliospheric Imagers

Nicholeen M. Viall[1] • Harlan E. Spence[2] • Angelos Vourlidas[3] • Russell Howard[3]
1. NASA Goddard Space Flight Center, Greenbelt, MD, USA,
   Nicholeen.M.Viall@nasa.gov
2. University of New Hampshire, Space Science Center, Durham, NH
3. Naval Research Laboratory, Solar Physics Branch, Washington, DC

**Abstract**

We present an analysis of small-scale, periodic, solar-wind density enhancements (length-scales as small as ≈ 1000 Mm) observed in images from the *Heliospheric Imager* (HI) aboard STEREO A. We discuss their possible relationship to periodic fluctuations of the proton density that have been identified at 1 AU using *in-situ* plasma measurements. Specifically, Viall, Kepko, and Spence (2008) examined 11 years of *in-situ* solar-wind density measurements at 1 AU and demonstrated that not only turbulent structures, but also non-turbulent periodic density structures exist in the solar wind with scale sizes of hundreds to one thousand Mm. In a subsequent paper, Viall, Spence, and Kasper (2009) analyzed the α to proton solar-wind abundance ratio measured during one such event of periodic density structures, demonstrating that the plasma behavior was highly suggestive that either temporally or spatially varying coronal source plasma created those density structures. Large periodic density structures observed at 1 AU, which were generated in the corona, can be observable in coronal and heliospheric white-light images if they possess sufficiently high density contrast. Indeed, we identify such periodic density structures as they enter the HI field of view and follow them as they advect with the solar wind through the images. The smaller periodic density structures that we identify in the images are comparable in size to the larger structures analyzed *in-situ* at 1 AU, yielding further evidence that periodic density enhancements are a consequence of coronal activity as the solar wind is formed.

## 1. Introduction

The ambient solar-wind plasma density at 1 AU is often composed of periodic density fluctuations on timescales of minutes to tens of minutes and with radial length-scales of ≈ 70 – 3000 Mm (Kepko, Spence, and Singer, 2002; Stephenson and Walker, 2002; Kepko and Spence, 2003). These density variations are inherently periodic, exhibiting power at discrete frequencies and wavelengths, identified through spectral analysis (Kepko, Spence, and Singer, 2002; Kepko and Spence, 2003; Viall, Kepko, and Spence, 2008; Viall, Kepko, and Spence, 2009). Through event studies, Kepko and Spence (2003) identified density fluctuations with discrete periodicities between $f$ = 0.1-3 mHz (three hours – five minutes). They showed that the lowest discrete frequencies ($f$ ≈ 0.1 mHz) corresponded to a radial length scale of ≈ 3200 Mm. Additionally, Kepko and Spence (2003) and Viall, Kepko, and Spence (2009) showed instances in which density fluctuations on time scales of hours have shorter periodicities embedded within them.

Although turbulence contributes to the fluctuations observed in the ambient solar-wind at 1 AU, Viall, Kepko, and Spence (2008) demonstrated that non-turbulent, periodic density structures also contribute significantly to fluctuations at 1 AU. Specifically, Viall, Kepko, and Spence (2008) divided 11 years of *in-situ* solar-wind density measurements at 1 AU into short (four – seven hour) data series and performed spectral analysis on each one. They identified all statistically significant radial wavelengths between ≈ 70 and 900 Mm through the application of two spectral tests: a narrow-band amplitude test, and a harmonic F-test (Mann and Lees, 1996; Thomson, 1982). They showed that periodic solar-wind number density structures occur more often at particular radial length-scales. For the slow solar-wind (<550 km s$^{-1}$), radial length scales that are most prevalent are $L ≈$ 73, 120, 136, and 180 Mm, and for the fast solar-wind those length scales are $L ≈$ 187, 270, and 400 Mm. Periodic density structures occurred in 70 – 80% of the slow solar-wind analyzed and 40 – 60% of the fast. Viall, Kepko, and Spence (2009) repeated the study in the time – frequency domain, investigating the apparent frequency of the structures in the rest frame of the spacecraft, analyzing between $f = 0.5 – 5$ mHz (30 – 3 minutes). Consistent with the previous studies, they found that particular discrete frequencies occur more often than others. The observation of particular radial length scales or frequencies occurring more often than others cannot be explained by theories of turbulence alone and requires some organizing physical process.

      Kepko, Spence, and Singer (2002) and Kepko and Spence (2003) suggested that a possible generation mechanism of periodic density structures is located in the solar corona and operates as the solar wind is formed. In this scenario, periodic density structures are entrained in the solar-wind flow shortly after the formation of that solar-wind. They advect out to 1 AU largely intact and they are observed in *in-situ* plasma measurements. Viall, Spence, and Kasper (2009) confirmed the solar corona as a likely source location of periodic density structures through the analysis of the α to proton solar-wind abundance ratio measurements during an event of periodic density structures. They presented an event in which the alpha density varies with the same periodicity as the protons, but is in antiphase with the protons. They confirm the significance of the periodicity utilizing the same spectral analysis methods of the previous studies. The observed solar-wind alternated between plasma composed of high proton density and low alpha density and plasma composed of low proton density and high alpha density in a periodic nature, and on a time scale of ≈30 minutes. This behavior strongly suggests that either temporally or spatially varying coronal source plasma created the periodic density structures.

      If at least some periodic density structures are in fact generated in the solar corona, then in principle we should be able to identify them (at least those large enough to be resolved by the imaging instruments) as they are released from the corona and entrained in the solar wind. Here, we present evidence of such periodic density structures advecting with the ambient solar-wind through images obtained by the imagers on the STEREO-A spacecraft. We examine nine days (14 January – 23 January 2008) of white light observations from 15 – 80 solar radii [$R_S$]. We identify quasi-periodic density structures in the SECCHI HI1 images that are comparable in size to the larger structures identified *in-situ* at 1 AU.

## 2. Data Analysis

      This analysis is different from the previous studies of periodic density structures which all employed *in-situ* plasma data measurements made at 1 AU. In this study, we

use, for the first time, remote imaging data. We analyze white light images of the solar wind taken with the *Heliospheric Imager* (Harrison, Davis, Eyles *et al*., 2008) on *Sun Earth Connection Coronal and Heliospheric Investigation* (SECCHI) (Howard, Moses, Vourlidas*et al.,* 2008) aboard the *Solar Terrestrial Relations Observatory Ahead* (STEREO-A) spacecraft (Kaiser, Kucera, Davila *et al.*, 2008). STEREO-A is on a heliocentric orbit in the Ecliptic, slightly inside of Earth's orbit and traveling slightly ahead of the Earth. Using images taken from 14 – 23 January 2008, we look for and identify periodic density structures advecting away from the Sun.

An important characteristic of the periodic density structures is that though they appear as periodicities in time in the rest frame of Earth and the spacecraft, they are not waves propagating through the solar wind (Kepko, Spence, and Singer, 2002; Kepko and Spence, 2003; Viall, Kepko, and Spence, 2008, 2009). The periodicities are due to spatial variation embedded in the solar wind, which advect passed the spacecraft. In fact, this distinction between a variation in time and a spatial variation that advects passed the observing spacecraft is difficult to discern unambiguously with single-point measurements. Thus, images allow for the observation of an entire train of periodic density structures at one time.

The inner *Heliospheric Imager* (HI1) on SECCHI has a 20° field of view, is pointed 14° away from the Sun, and its optical axis is aligned with the ecliptic plane. The resultant field of view of HI1 in solar elongation angle is from 4 – 24°, which corresponds to a projected heliocentric distance of 15 – 84 $R_S$ from the Sun perpendicular to the line of sight. The HI1 images have 1024 x 1024 pixels, and a platescale of 72 arcsecs per pixel. Images are obtained every 40 minutes; each single image is a sum of many short exposures, with a total integration time of 30 minutes.

The emission in the HI1 images arises from the Thomson scattering of the photospheric light by the electrons in the corona and solar wind (Billings, 1966). The brightness in the images is proportional to electron density. This scattered emission has a maximum intensity where the angle between the incident radiation on the electron and the observer's line of sight through the electron is 90°. For solar wind electrons, this occurs on the 'Thomson Sphere', where the angle between the center of the Sun, the emitting solar-wind electrons and the observer is equal to 90°, as described in Vourlidas and Howard (2006). The diameter of the Thomson Sphere is equal to the distance between the observer and the center of the Sun, and centered in between the Sun and the observer. For the analysis that we present in this paper, we examine primarily the inner half of the SECCHI HI1 images (*i.e.* the half nearest the Sun). We focus in particular on the portion of the image with a field of view of about 4 – 14° away from the Sun. At these small solar elongation angles, the lines of sight of each pixel are approximately parallel and the Thomson Sphere can be approximated as a plane coincident with the plane of the solar limb (Vourlidas and Howard*,* 2006).

The image cadence of the HI1 data limits our ability to resolve the entire range of periodic density enhancements of the size scales of the aforementioned studies on periodic solar-wind density structures. However, both spatially (the wavelength) and temporally (the frequency) there is overlap in the analyzed spectral range. The temporal cadence, which is relevant for analysis of density fluctuations in a given pixel or group of pixels as a function of time, results in a Nyquist frequency of 0.2 mHz, or 80 minutes. The periodicities that can be analyzed given this resolution are commensurate with the longest events studied in Kepko and Spence (2003) and Viall, Kepko, and Spence (2009). In particular, Viall, Kepko, and Spence (2009) identified a periodicity around $f$ = 0.2 mHz in their event study, and Kepko and Spence (2003) identified periodicities as low as $f$ = 0.1 mHz (≈ three hours) in their event studies.

While the HI platescale is 72 arcsecs per pixel, the instrument modulation transfer function (MTF) extends to ≈ 3 pixels, so the effective resolution is ≈ 2.5 arcmin. In January 2008, STEREO A was located at a radial distance of 0.9673 AU from the Sun, therefore each pixel corresponds to a distance of 48 Mm perpendicular to the line of sight, using the plane-of-sky assumption. Due to the three-pixel effective resolution (three pixels = 144 Mm), periodic structures as small as 144 Mm can be resolved. The radial length-scale of the structures examined by Kepko and Spence (2003) and Viall, Kepko, and Spence (2008) were in the range of 70 – 3000 Mm, so the spatial resolution of HI1 is just adequate to resolve the structures of interest.

Although the spatial resolution is ≈ 300 Mm, the 30-minute integration time per image will effectively decrease the spatial resolution, due to velocity smearing as the structures propagate across the field of view during integration time. As an example, density enhancements moving at a typical slow solar-wind speed (at ≈ 15 – 20 $R_S$) of 300 km s$^{-1}$ perpendicular to the line of sight will travel 540 Mm, or about 11 pixels, during the 30-minute integration time of a single image. Such density enhancements contribute intensity to all of those 11 pixels, thus limiting our effective spatial resolution. In Figure 1, we plot this pixel smear as a function of solar-wind speed. For the solar-wind speeds relevant here, 200-350 km s$^{-1}$, the distance traveled over 30 minutes is 323 – 603 Mm. Therefore, periodicities as small as ≈ 650 – 1200 Mm can still be identified, even when accounting for smear caused by the solar wind velocity. In the statistical analysis of Viall, Kepko, and Spence (2008), periodic radial length-scales as large as ≈ 700 – 1000 Mm were often identified. Specifically, they identified that an occurrence enhancement at 700 – 800 Mm persisted at the > 2 σ level for six of the nine three-year occurrence distributions. Large size scales such as these and the few-thousand-Mm structures identified in Kepko and Spence (2003) should be resolvable in the SECCHI HI1 data.

In order to identify more easily the small-scale (small relative to other features in the SECCHI HI1 images) density enhancements, we constructed ratio images as follows. First, we calculated the median value of each pixel over the nine days of data we examine here, 14 – 23 January 2008. The ratio images are created by dividing each of our images by the median image. The result is that features that are constant over those nine days are removed, while features that evolve or move on time scales shorter than nine days remain intact and are more easily discerned. This processing allows in particular for easier identification of advecting density fluctuations with length scales we are seeking: 1 000- 10 000 Mm.

In the next section, we present an event analysis of these HI1A ratio images. We identify periodic advecting density structures in the HI1 field of view, and follow the train of enhancements as they advect away from the Sun. We discern large-scale periodicities as well as smaller periodic structures that persist. The smallest resolvable periodic structures that we identify are commensurate with the largest structures identified in the previously described *in-situ* studies. We confirm that density structures do not propagate relative to the solar wind; rather they are entrained in the ambient solar-wind. Lastly, we calculate the time series of a slit of pixels which the density structures pass through and use spectral analysis to confirm the presence of significant discrete periodicities. Structures such as those we examine here are likely a source of periodic density variations observed *in-situ* at 1AU.

2.1 Event 1

Within the nine days of data analyzed, we focus on two events that contain many periodic density enhancements. Event 1 is approximately one day long, beginning at 20:09 on 19 January 2009, and lasting until 15:29 on 20 January 2008. We examine the time series of the processed SECCHI images, identify structures as they enter the SECCHI HI1A field of view, and follow them as they advect outward in subsequent images. In Figure 2 we show a SECCHI HI1A image during event 1 in which we have identified periodic density structures. This image was taken on 20 January 2008, at 07:29:01 UT. The color bar on the left of the image indicates the pixel ratio value. Although the full SECCHI HI1 image is 1024 x 1024 pixels, here we show only the ecliptic-centered 512 pixels and the sunward 512 pixels (one quarter of the total HI1 image). In this way, we focus on the part of the image that contains the periodic structures as well as focus on the region closest to the Sun, where the HI1A signal to noise is best. In all of the images that we present in this paper, the Sun is to the right of the field of view and the solar wind travels from right to left in time. In the HI1A images, the motion of the stars is from left to right in time (nearly opposite the motion of the solar wind in this region). The horizontal axis is the radial distance from the Sun and both the radial and vertical distances are given in units of pixels (recall that each pixel corresponds to ≈ 48 Mm perpendicular to the line of sight).

There are numerous variations in intensity evident throughout this image on both large and small scales, and in both the vertical and radial direction. We highlight one of many regions in the image that contains small-scale, periodic intensity variations in the radial direction with a white box. We also indicate a constant radial slice with white lines, which contains periodic density structures. We include two grayscale movies in the supplementary material of the HI1A ratio images over these nine days. The first movie shows the same 512 x 512 pixel array as in Figure 2 (and Figure 7) of the structures. In the second movie, we show the same nine days, but focus on only a 100 x 100 pixel array. The yellow- boxed region of Figure 2 defines that 100 x 100 pixel field of view. Clear variations that advect with the solar wind can be identified as they pass through the field of view.

In Figure 3, we show a time sequence of SECCHI HI1A images that contain the periodic density structures identified in Figure 2. The images are composed of the white boxed region of pixels indicated in Figure 2. The images advance in time from the bottom to top of the column, and from the right to left column. We show every fourth image, starting with the image taken at 20:49 on 19 January 2008. Note that the white-boxed region of the image displayed in Figure 2 is displayed in the left column, bottom image (highlighted with green box around it) in Figure 3. The color scale is the same color scale used in Figure 1, shown in the upper middle of Figure 3. We list the date and time that the image was taken to the right of each image. As in Figure 2, the radial and vertical distances are given in units of pixels.

Whenever a density enhancement enters the field of view, we identify it visually with a letter (beginning with "A", and following in alphabetical order). We follow the density structure in the time sequence of images, indicating its location with an arrow and its letter identification. We identify a train of such enhancements, A through M, in the images presented for event 1. Intensity enhancements that are in fact due to stars will move from left to right in time (towards the Sun), and therefore can be easily differentiated from small solar-wind enhancements which will move from right to left with time (away from the Sun).

As the periodic density structures advect outwards, their individual identity is more difficult to discern, and often an envelope structure containing a few smaller structures becomes the feature most evident in the images. When this occurs, we maintain the labels we have given the individual structures, but assign the overall envelope structure only one arrow. The apparent loss of identity of individual structures as they advect outwards is likely the result of a lower signal to noise ratio further from the Sun. It also can occur when viewing structures edge-on that expand azimuthally, a point we return to in the discussion section.

Evident are large-scale periodic structures that enter the field of view, and advect with the solar wind. For example, image 20:49 on 19 January 2008 contains a 75-pixel (≈3600 Mm) periodicity indicated with circles. We identify another large-scale periodicity in image 10:09 on 20 January 2008 with three circles. The first circle encompasses "ABCDE", the second encompasses "FGHI", and the third encompasses "KL". The circles are spaced at about 150 pixel intervals (7000 Mm). Examining a previous image (07:29 on 20 January 2008) reveals that the large-scale structure has embedded within it a smaller, 25 pixel (1200 Mm) periodicity: G, H, I, J, and K.

In order to gain a quantitative assessment of the presence of small-scale periodic structures advecting with the solar wind beyond this qualitative visual assessment, we compute radial intensity profiles of each image (including those in between those shown here). We compute each radial profile by taking the median value of a vertical slice of pixels, one pixel wide, across the radial cut indicated in Figures 2 and 3 bounded by the two white lines. Those white lines indicated in Figure 2 define the height of the vertical slice with radial distance. We take the median value of the vertical slice of pixels, as the median will be the most insensitive to point sources (*i.e*., stars). For 30 images, beginning with the image taken on 20:09 on 19 January 2008, we compute these radial profiles of the periodic density structures, and plot them in Figures 4a and 4b. We offset each subsequent radial profile by 0.002 in *y*. The tick marks on the *y*-axis are equal to that offset of 0.002.

We list time to the left of each radial profile; time increases from bottom to top and right to left, as in the image sequences. The time labels of those profiles that have a corresponding image shown in Figure 3 are underlined. On top of the median pixel values (grey) we plot the six-point running average (black) to further smooth point enhancements from stars. We identify solar-wind enhancements as they enter the field of view, label them such that they correspond to those we identified in Figure 3, and follow them in time. This form of plotting these data collapses the two dimensional images into a quantitative one-dimensional plot. In this way, smaller-scale density structures can be more easily identified and differentiated from false enhancements due to stars.

Those enhancements that persist and advect with the solar wind (*i.e.* those that are part of a train of discrete density enhancements) are evident and marked with letters corresponding to their identification letter in the images. We identify enhancements as individual structures from multiple criteria. First, each structure that we identify in the radial profiles correlates with clear, distinct enhancements in the images shown in Figure 3. In the radial profiles, we require that the structure is a distinct enhancement through at least three radial profiles, and advects with the overall train of structures for us to identify the structure as a separate enhancement with its own label. We do not require that every individual structure persist in multiple images in Figure 3, as we only show every fourth image; however, even the small structures often persist in multiple images shown in Figure 3.

In these radial profiles, the large-scale periodicities that we identified with circles in Figure 3 are evident as enhancements, persist through many radial profiles, and advect together as we would expect enhancements in the ambient solar-wind to do. For example, the large, 150-pixel structures identified in image 10:09 on 20 January 2008 can be seen in the corresponding radial profile as three broad peaks spaced about 150 pixels apart. Likewise, the shorter size-scale enhancements in image 07:29 on 20 January 2008 (enhancements G, H, I, J and K), are evident in the corresponding radial profile. In the radial profiles, it is clear that even these small enhancements persist in subsequent time steps, advecting with the ambient solar-wind. Looking further back in time, we see that they enter the field of view as distinct enhancements with depletions in between them. As in Figure 3, as the structures advect further out their individual identity is lost, and only the large-scale envelope structure is discernable. This shorter 1200 Mm periodicity is commensurate with the larger size scales observed at 1 AU.

We estimate the velocity of the structures, using the assumption that they are moving perpendicular to the line of sight, and the plane of sky assumption. One enhancement that persists for a long period of time and is clearly differentiated from other enhancements is "H". The leading edge of H enters the SECCHI HI1A field of view at 01:29 on 20 January. The approximate centroid (defined as the peak relative intensity in the radial profiles) of feature H travels 112 pixels (to pixel 400) by 07:29 UT on that same day, giving an average velocity of 250 km s$^{-1}$ perpendicular to the line of sight. It travels another 100 pixels by 12:09 UT on that day, which is an average velocity over those five hours of 285 km s$^{-1}$. At 15 solar radii away from the Sun (the inner edge of the SECCHI HI1 field of view), the solar wind is not fully accelerated to the velocities observed at 1 AU, however it has completed most of its acceleration, and is beyond the Alfvén and sonic transition points. H is traveling at a slow solar-wind speed, and accelerates some, consistent with the behavior expected of slow solar-wind between 15 and 50 $R_S$. This estimate confirms that the line of sight and plane of sky approximations are not unreasonable in estimating the size scales of these structures. Further, this is consistent with the concept that these periodic density enhancements are entrained features of the ambient solar-wind.

Next, we investigate the time series of a slit of pixels at the inner edge of the HI1 field of view. This is analogous to what a spacecraft measuring plasma *in-situ* might observe if it were located in the solar wind plasma that those pixels are imaging. Figure 5 shows the median value of vertical pixels 255:270, at radial pixel 505, as a function of time. The periodic train of structures all pass through these pixels as they enter the HI1 field of view and advect outwards. We identify when a given structure passed pixel 505 with the corresponding letter, using the same letters used in Figures 3 and 4. We identify visually in the time series a significant five-hour periodicity with red dashes as well as a three-hour periodicity with green dashes, and a 100-minute periodicity with cyan dashes to guide the eye, but follow that with a quantitative analysis.

We perform spectral analysis and test the significance of the identified periodicities in the time series following the methods of Man and Lees (1996) and Thomson (1982), as implemented in Viall, Kepko, and Spence (2008) and Viall, Kepko, and Spence (2009). In Figure 6 we plot the spectral estimate (thick black) of the time series shown in Figure 5 calculated using Thomson's multitaper method (Thomson, 1982). There are spectral peaks at five hours (red dash), three hours (green dash) and 100 minutes (cyan dash), which are the periodicities that we identified by eye in the time series. We perform two independent tests for peak significance. First, we perform a narrow-band amplitude test, which tests the significance of a spectral peak relative to the background spectra (Mann and Lees, 1996; Thomson, 1982). We plot the background spectra (thin black line) along with the 95% and 90% confidence levels (grey) in Figure 6. Second, we perform a harmonic F-test, which tests for phase coherent signals (not shown) (Mann and Lees, 1996; Thomson, 1982). As in our earlier studies of periodic solar-wind density structures, we consider a spectral peak to be significant only if it passes both tests simultaneously at the chosen confidence threshold. The five-hour spectral peak passes both the harmonic F-test and the narrow-band test at the 95% threshold simultaneously, and is underlined. The three-hour and 100-minute periodicities pass both tests at the 90% confidence threshold. Periodic density structures at both of these frequencies have been observed at 1 AU.

We test the HI1 A images for harmonic instrumental artifacts that may introduce power into a spectral estimate, such as the one shown in Figure 6. We construct time series of every pixel in the entire HI1 A image covering the nine-day interval studied here. We performed spectral analysis on each of these time series using the methods described above. We identify a periodicity that is highly phase-coherent (it has a high harmonic F-test value) and has significant discrete power at $f = 0.139$ mHz (two hours) in almost every pixel time series, irrespective of whether the pixels contained coronal signal or not. Other spectral peaks, such as the five-hour, three-hour and 100-minute peaks that we identify here, are only present in pixels that contained the density structures. In the spectra calculated from the time series shown in Figure 4, the two-hr signal only passes both tests at the 90% threshold, as it has little power. In general, we found that in the corners and outer edge of the SECCHI HI1 image, where the brightness from the solar wind is very low, the two-hr signal has a much higher significance level in both spectral tests. In contrast, in pixels where there is a substantial signal from the solar wind (e.g. towards the center of the image and in the sunward portion of the image), the significance of the two-hr signal is suppressed. The origin of this signal is unknown and is currently being investigated by the SECCHI instrument team. Therefore, we choose the conservative approach and treat the $f = 0.139$ mHz signal in this event as a suspect signal, and not due to true variation in the solar wind plasma. This highlights the importance for these images of correlating spectral peaks with true enhancements in the time series of the pixel slit, and further correlating the enhancements in the time series of the pixel slit with structure in the images that advect with the solar wind, as we have done here.

2.2 Event 2

We identify a second, longer event a few days earlier and repeat the analysis of periodic density enhancements in this sequence of images, as we did for Event 1. The second event that we identified begins at 00:09 on 14 January 2008, and lasts until 20:49 on 16 January 2008. In Figure 7 we present a representative image in the same format as Figure 2 for this event. This image was taken at 14:49 on 16 January 2008, and is composed of the same 512 x 512 pixel field of view that we used in Figure 2. Mercury is visible in the left side of Figure 7. We identify one region containing periodic enhancements with a white box and white lines indicating a radial slice of constant angle that we use to calculate the radial profiles (Figure 9). To the left of the image we show the color bar of relative pixel intensity.

In Figures 8a – d, we show every fourth image in a sequence of images spanning event 2, in the same format as Figure 3. Time increases upwards, and from right to left column. As before, advecting solar-wind density enhancements move left with time and stars move to the right with time. We identify and label with letters periodic density enhancements as they enter the field of view, and follow them as they advect away from the Sun with the solar wind. We identify structures A through T in event 2. The color scale is shown between the columns in the image sequences and are the same as in Figure 7, with the exception of the image sequence shown in 8b. For illustration purposes, we chose a slightly different color bar for four of the images in image sequence 8b (04:49 to 12:49 on 15 January 2008), shown to the right of image 04:49 on 15 January 2008.

We highlight some of the periodic structures within the event with circles. In image 04:49, 01/14/08, we identify three enhancements (B, C, and D) which have ≈ 30-pixel (≈1500 Mm) spacing. In image 10:09, 14 January 2008 we identify four structures, A, BC, D, and E, which occur at a ≈100 pixels (≈5000 Mm) spacing. In image 02:09 on 15 January 2008, structures F, G, H, and I occur with ≈ 50 pixel spacing. That train of enhancements continues, and in image 18:09 on 15 January 2008 we identify K, L, M and the beginning of N, which are also about 50 pixels apart.

In image 04:49 on 16 January 2008, M, N, and O occur with a 60-pixel spacing, followed by P, Q, and R in image 12:49, 16 January 2008 which also occur with a ≈ 60-pixel spacing. In the same image, the large structures NO, P, R, and S occur about 125 pixels apart. In the context of the previous images, it is evident that NO is in fact composed of two structures, and that in between P and R was a clear structure "Q". In other words, there is smaller, substructure within the more apparent large-scale 125-pixel structure. Finally, in image 15:29, 16 January 2008, we see O, P, Q, R, and S, which are the same structures visible in Figure 7.

In Figure 9, we quantify the radial profiles of the periodic density structures identified in Figure 8 for event 2. Using the same letters as in Figure 8, we identify structures as they become enhancements in the radial profiles (when they enter the field of view) and follow them in each subsequent radial profile. The time and date of the image from which the radial profile is calculated is shown to the left; those that have a corresponding image in Figure 8 are underlined. We begin with the radial slice computed from the image taken at 00:09 on 14 January 2008, and show radial slices for every image through 20:49 on 16 January 2008. As in event 1, the radial profiles allow for the distinction of smaller structures embedded in larger enhancements.

Only those enhancements that are seen clearly in Figure 8 are intensity peaks in multiple radial profiles, and advect in concert with the overall train of solar-wind structures, are identified as individual structures. We identify structures A through T in the radial profiles, and follow them in time. We confirm quantitatively that these structures which we identify visually in Figure 8 are clear periodic enhancements in the radial profiles. For example, the image taken 10:09 on 14 January 2008 (Figure 8a) has the corresponding radial slice plotted in Figure 9a. Consistent with the four large enhancements identified with circles in 8a, there is a large enhancement, A, centered near pixel 200, an enhancement near 300 labeled BC, an enhancement D at about 375, and a very large enhancement E at about 475. As we identified in the image, the enhancements are ≈100 pixels apart. As another example, the radial slice from image 02:09 on 15 January 2008 (Figure 9b), indeed shows that F, G, H, and I are about 50 pixels apart. In image 18:09 on 15 January 2008, L, M and N are enhancements located about 50 pixels apart. In this radial profile, K is lost in front of a star, however examining prior radial profiles (*e.g*. 14:49, 15 January 2008, Figure 9c) and subsequent radial profiles (*e.g*. 20:09, 15 January 2008) we see K emerge again as an enhancement, adding to the train of periodic density structures spaced at ≈50 pixel intervals. This illustrates the utility of the radial profiles for confirming the existence and persistence of smaller structures in the solar wind. Finally, the structure O, P, Q, R, S seen in Figure 7 and 8d is seen in 9e at 14:49. The train of enhancements advect together in time.

In Figure 10, we plot the time series of a slit of pixels at the inner edge of the field of view, through which all of the identified structures passed. We compute the median value of vertical pixels 260 – 280 at radial pixel 505, as a function of time. We mark the time at which each individual structure crosses pixel 505 in the time series with the letter of the structure. It is clear in the time series that each enhancement we identified as a advecting structure in the image sequences and in the radial profiles indeed passes through that slit of pixels, yielding variations in the time series of those pixel values. We identify visually and denote a nine-hour periodicity (red), a 2.5-hour periodicity (cyan) and a four-hour periodicity (green) in the time series for reference.

We confirm quantitatively that these periodicities are statistically significant using the spectral methods described and implemented for event 1. We show the spectral estimate (thick black), with the background spectra (thin black) as well as the 95% and 90% narrow-band confidence thresholds (grey) in Figure 11. As always, we require that spectral peaks must pass both the narrow-band test and the harmonic F-test simultaneously at a given confidence threshold. With this requirement, the four-hour periodicity is significant at the 95% confidence threshold. The 9 hour and 2.5-hour periodicities are significant at the 90% threshold. The artificial signal is present again at the 90% threshold, but again it is discounted.

Event 2 is longer than event 1, lasting for a full three days. In the three-day image sequence, it is clear that trains of periodic density structures can persist for many days. Furthermore, even over the three days, there is coherence to the periodicities that are present; in this event, the four-hour periodicity is strong for a significant portion of the three-day event. As in the first event, the smaller scale periodicities identified here (e.g. $L$=1500 Mm and $t$ = 2.5 hour) are similar to those identified in the 1 AU studies.

## 3. Discussion

In the events described above, the distinction of individual plasma enhancements is difficult to discern across the entire HI1 image. The smaller ones specifically tend to persist for a few images before they appear to fade and coalesce with preceding and following enhancements, forming a large 'envelope' structure. These larger scale structures generally persist through the field of view presented in Figures 3 and 9.

As we pointed out in the discussion of Figure 3, a lower signal-to-noise level occurs in the HI1 images with increasing distance from the Sun, which will cause structures to fade into the background noise as they advect away from the Sun. Additionally, when viewed edge on, even structures that maintain their identity as distinct physical features may appear to coalesce in the HI1 field of view as they advect outward due to azimuthal expansion. We expect that there will be azimuthal expansion of periodic density structures with radial distance from the Sun, as the solar wind is space filling. We illustrate two scenarios in Figure 12 in which there are separate, periodic enhancements entering the field of view, which maintain their radial size and separation but will appear to merge into one large structure due entirely to azimuthal expansion. In both scenarios, the structures are in the Ecliptic, labeled with letters as we did in the images, and the lines of sight from SECCHI HI1 are edge on.

In Figure 12a, we illustrate a scenario in which the normal of the fronts of the periodic density structures are aligned with the solar wind flow direction, but the line of sight from SECCHI HI1 is not perpendicular to the solar wind flow in which they are entrained. In Figure 12b we illustrate a scenario in which the periodic density structures are advecting perpendicular to the line of sight from SECCHI HI1, however the normal of the fronts are inclined relative to the flow direction. Both scenarios result in distinct periodic structures that can be distinguished as they enter the field of view (lines of sight 9, 10, and 11), but gradually appear in projection to coalesce as they advect outwards (lines of sight $1 - 5$), due entirely from azimuthal expansion and not due to true physical coalescence.

The periodic structures that we examine in this paper, and therefore the periodic density structures observed *in-situ* at 1 AU, could be related to helmet-streamer plasmoids which have been studied extensively (*e.g.* Sheeley, Wang, Hawley *et al*., 1997; Wang, Sheeley, Walters *et al.*, 1998; Sheeley, Warren, and Wang, 2007; Harrison, Davies, Rouillard, Savani, Davies *et al*., 2009; Sheeley, Lee, Casto *et al*., 2009; Rouillard, Savani, Davies *et al*., 2009; Rouillard, Davies, Lavraud *et al*., 2010a; Rouillard, Lavraud, Davies *et al*., 2010b). Both phenomena are thought to contribute plasma structures to the ambient slow solar-wind. However, the most important distinction between the helmet-streamer plasmoids previously studied and these event studies is the inherent periodic nature to the structures we analyze here. Additionally, the helmet-streamer plasmoids are on the larger end of the size scales that we study here and that were examined at 1 AU. For example, Sheeley, Wang, Hawley *et al*. (1997) found that the size scales of helmet streamer plasmoids are ≈ 3000 Mm by the time they reach 15 $R_S$. In this study and in the studies at 1 AU, periodic density structures as large as 3000 Mm were identified, however we identified scales as small as 1000 Mm in this event study and as small as 75 Mm were identified in the studies at 1 AU.

Further, Sheeley, Lee, Casto *et al*. (2009), Sheeley, Warren, and Wang (2007) and Wang, Sheeley, Walters *et al.* (1998) find that typically the helmet-streamer plasmoids occur at a rate of four per day (one every six hours) during solar minimum, whereas we have identified as many as 16 in a single day (event 1). Although we identified large-scale structures with periodicities on these sorts of time scales, we also identified small structures embedded within the larger envelope structure with time scales down to 100-minutes. A parallel therefore can be drawn between the large-scale envelope structures examined here and the helmet-streamer plasmoids, while the smaller, embedded substructure have a size scale in common with the density structures studied at 1 AU. Likewise, Kepko, Spence, and Singer (2002), Kepko and Spence (2003) and Viall, Kepko, and Spence (2008 and 2009b), found that often the largest (≈ 1000 Mm) *in-situ* periodic structures had smaller periodic structures embedded within them.

A further distinction between the helmet-streamer blobs and the periodic density structures studied *in-situ* at 1 AU is that the streamer blobs are a plasma sheet phenomenon, and they are not thought to comprise the entirety of the slow solar-wind (Wang, Sheeley, Socker *et al.,* 2000). In contrast, Viall, Kepko, and Spence (2008) showed that periodic density structures occurred in as much as 80% of the slow solar-wind, occurred during the entirety of the solar cycle, and even occurred in fast (>550 km s$^{-1}$) solar-wind, albeit with a lower occurrence rate.

**4. Summary and Conclusion**

We perform an exploratory study of periodic density structures in SECCHI HI1A images. This is the first time that density structures that are inherently periodic have been identified and analyzed in images of the solar wind. We present two events in which we look for and identify periodic density structures in the SEHCCI HI1 images advecting from the Sun with the ambient solar-wind. The structures most easily observed in these SECCHI HI1A images are generally larger than those that were focused on in the studies *in-situ* at 1 AU (Kepko, Spence, and Singer, 2002; Kepko and Spence, 2003; Viall, Kepko, and Spence, 2008, 2009; Viall, Spence, and Kasper, 2009). However, the smaller structures (≈1000 Mm) identified in the images are commensurate in spatial and temporal scales with the largest scales that were analyzed in the study of periodic density structures at 1 AU. This provides further evidence that at least some periodic density structures observed *in-situ* at 1 AU are generated somewhere in the solar corona as the solar wind is formed.

We confirm that periodic density structures in the solar wind beyond ≈ 15 $R_S$ are truly advecting spatial structures rather than a time-dependent phenomena. We demonstrate that as the train of periodic density enhancements advects past a single location, the time series at that location exhibits a temporal periodicity. We conclude that periodic density structures are injected into the solar wind somewhere before ≈ 15 $R_S$, and they advect with the ambient slow solar wind to 1 AU where they are observed with *in-situ* plasma measurements. Though we have confirmed that the corona can be a source of periodic density structures, further research relating these structures to specific source locations in the solar corona is needed and is underway. Finally, the STEREO – Sun – Earth angle has been increasing since the STEREO spacecraft were placed in their final orbits. Soon, the STEREO – Sun – Earth angle will be near 90°, at which time density structures that are observed in the inner edge of HI1 (*i.e.* low elongation angles) will be on an Earthward trajectory. This configuration is well suited for a direct comparison between SECHHI HI observations and *in-situ* measurements made near Earth, and we plan to pursue such studies at the first opportunity.

**Acknowledgements**: We thank the STEREO/SECCHI team for the use of these excellent data. The STEREO/SECCHI data used here are produced by an international consortium of the Naval Research Laboratory (USA), Lockheed Martin Solar and Astrophysics Lab (USA), NASA Goddard Space Flight Center (USA) Rutherford Appleton Laboratory (UK), University of Birmingham (UK), Max-Planck-Institut für Sonnensystemforschung(Germany), Centre Spatiale de Liège (Belgium), Institut d'Optique Théorique et Appliqueé (France), Institut d'Astrophysique Spatiale (France). This research was supported by NASA Grant No. NNG05GK65G, and an appointment to the NASA Postdoctoral Program at the Goddard Space Flight Center, administered by Oak Ridge Associated Universities through a contract with NASA.

**References**

Billings, D.E.: 1966, *A guide to the solar corona*, New York: Academic Pres.

Harrison, R.A., Davis, C.J., Eyles, C.J., Bewsher, D., Crothers, S., Davies, J.A., Howard, R.A., Moses, D.J., Socker, D.G., Halain, J.-P., Defise, J.-M., Mazy, E., Rochus, P.,Webb, D.F., Simnett, G.M.: 2008, *Solar Phys*. **247**, 171.

Harrison, R.A., Davies, J. A., Rouillard, A. P., Davis, C. J., Eyles, C. J., Bewsher, D., Crothers, S. R., Howard, R. A., Sheeley, N. R., Vourlidas, A., Webb, D. F., Brown, D. S.; Dorrian, G. D.: 2009, *Solar Phys.* **256**, 219 .

Howard, R.A., Moses, J. D., Vourlidas, A., Newmark, J. S., Socker, D. G., Plunkett, S. P., Korendyke, C. M., Cook, J. W., Hurley, A., Davila, J. M. et al.,: 2008, *Space Science Rev*. **136**, 67.

Kaiser, M. L., Kucera, T. A., Davila, J. M., St. Cyr, O. C., Guhathakurta, M., & Christian, E.: 2008, *Space Science Rev.* **136**, 5.

Kepko, L., Spence, H. E.: 2003, *J. Geophys. Res*. **108**, 1257.

Kepko, L., Spence, H. E., Singer, H. J.: 2002*, Geophys. Res. Lett*. **29**, 080000.

Mann, M., Lees, J. M.: 1996, *Clim. Change. 33*, 409.

Rouillard, A. P., Savani, N. P., Davies, J. A., Lavraud, B., Forsyth, R. J., Morley, S. K., Opitz, A., Sheeley, N. R., Burlaga, L. F., Sauvaud, J.-A., Simunac, K. D. C., Luhmann, J. G., Galvin, A. B., Crothers, S. R., Davis, C. J., Harrison, R. A., Lockwood, M., Eyles, C. J., Bewsher, D., Brown, D. S., 2009, *Solar Phys.* **256**, 307.

Rouillard, A. P., Davies, J. A., Lavraud, B., Forsyth, R. J., Savani, N. P., Bewsher, D., Brown, D. S., Sheeley, N. R., Davis, C. J., Harrison, R. A., Howard, R. A., Vourlidas, A., Lockwood, M., Crothers, S. R., Eyles, C. J.: 2010, *J. Geophys. Res.* **115**, 4103.

Rouillard, A. P., Lavraud, B., Davies, J. A., Savani, N. P., Burlaga, L. F., Forsyth, R. J., Sauvaud, J.-A., Opitz, A., Lockwood, M., Luhmann, J. G., Simunac, K. D. C., Galvin, A. B., Davis, C. J., Harrison, R. A.: 2010, *J. Geophys. Res*. **115**, 4104.

Thomson, D. J.: 1982, IEEE **70**, 1055.

Sheeley, N. R. Jr., Warren, H. P., Wang, Y.-M.: 2007, *Astrophys. J.* **671**, 926.

Sheeley, N.R. Jr., Wang, Y., Hawley, S. H., Brueckner, G. E., Dere, K. P., Howard, R. A., Koomen, M. J., Korendyke, C. M., Michels, D. J., Paswaters, S. E., Socker, D. G., St. Cyr, O. C., Wang, D., Lamy, P. L., Llebaria, A., Schwenn, R., Simnett, G. M., Plunkett, S., Biesecker, D. A.: 1997, *Astrophys. J*. **484**, 472.

Sheeley, N. R. Jr., Lee, D. D.-H., Casto, K. P., Wang, Y.-M., Rich, N. B.: 2009, *Astrophys. J.* **694**, 1471.
Stephenson, J. A. E., Walker, A.D.M.: 2002, *Geophys. Res. Lett*. **29(9)**, 1297, doi:10.1029/2001GL014291.
Viall, N. M., Kepko, L., Spence, H.E.: 2008, *J. Geophys. Res*. **113**, A07101, doi:10.1029/2007JA012881.
Viall, N. M., Kepko, L., Spence, H.E.: 2009, *J. Geophys. Res*. **114**, A01201, doi:10.1029/2008JA013334.
Viall, N. M., Spence, H.E., Kasper, J.: 2009, *Geophys. Res. Lett*. **36**, L23102, doi:10.1029/2009GL041191.
Vourlidas, A., Howard, R. A.: 2006, *Astrophys. J*. **642**, 1216.
Wang, Y.-M, Sheeley, N.R. Jr., Walters, J. H., Brueckner, G. E., Howard, R. A., Michels, D. J., Lamy, P. L., Schwenn, R., Simnett, G. M: 1998, *Astrophys. J. Lett.* **498** L165.

Wang, Y.-M., Sheeley, N. R. Jr., Socker, D. G., Howard, R.A., Rich, N. B.: 2000, *J. Geophys. Res.* **105**, A11.

**Figure 1** Pixel smear over 30-minute image integration time as a function of solar-wind speed perpendicular to the line of sight.

**Figure 2** SECCHI HI1A image taken on 20 January 2008, at 07:29:01 UT; event 1. The Sun is to the right of the image. The platescale is 72 arcsecs per pixel. A color bar indicating the relative pixel intensity is shown at the left of the image. The white box indicates portion of image shown in Figure 3; white radial cut indicates slice used for Figures 4a and 4b.

**Figure 3** Every fourth image in a sequence of SECCHI HIA images taken 20:49, 19 January 2008 through 15:29, 20 January 2008. Time increases up and from right to left. Letters and arrows identify plasma enhancements. Circles highlight periodic density enhancements. The date and time of image are shown to the right of each image. Color bar indicates relative pixel intensity.

**Figure 4** Radial profiles of the radial cut encompassing periodic density structures in the SECCHI HI1 images, 19 – 20 January 2008. Time increases upwards and from right to left; each profile is offset in *y* by .002 relative intensity units. Letters identify plasma enhancements as in Figure 3. Time and date of image is shown to the left; those that have a corresponding image in Figure 3 are underlined. a) Slices from 20:09 on 19 January 2008 until 08:49 on 20 January 2008. b) Slices from 09:29 until 15:29 on 20 January 2008.

**Figure 5** Time series of the median of vertical pixels 255:270, at radial pixel 505, 19 – 20 January 2008. We identify a five-hour (red), three-hour (green) and 100-minute (cyan) periodicity with dashes. Letters identify individual plasma enhancements, as identified in Figures 3 and 4.

**Figure 6** Spectral estimate (thick black) of time series shown in Figure 5. Ticks identify significant spectral enhancements in the same colors as Figure 5. The spectral background (thin black) and narrow-band significance thresholds (grey, 95% and 90% confidence thresholds) are shown.

**Figure 7** SECCHI HI1A image taken on 16 January 2008, at 14:49:01 UT; event 2. The Sun is to the right of the image and Mercury is in the left portion of the image. Platescale = 72 arcsecs per pixel. A color bar indicating the relative pixel intensity is shown at the left of the image. White box indicates portion of image shown in Figure 8a-d; white radial cut indicates slice used for Figures 11a – f.

**Figure 8** Event 2. Every fourth image in a sequence of SECCHI images taken 14 – 16 January 2008. Time increases up and from right to left. Letters and arrows identify plasma enhancements. Circles highlight periodic density enhancements. Date and time of image is shown to the right of each image. Color bar indicates relative pixel intensity. a) Images from 00:09 until 18:09 on 14 January 2008. b) 20:49 on 14 January 2008 until 15:29 on 15 January 2008. Note color bar change. c) 18:09 on 15 January 2008 until 12:49 on 16 January 2008. d) 15:29 until 20:49 on 16 January 2008.

**Figure 9** Radial profiles of the radial cut encompassing periodic density structures in the SECCHI HI1 images, 14 – 16 January 2008. Time increases upwards and from right to left; each profile is offset in *y* by .002 relative intensity units. Letters identify plasma enhancements as in Figure 8. Time and date of image is shown to the left; those that have a corresponding image in Figure 8 are underlined. a) Slices from 00:09 until 12:49 on 14 January, 2008. b) Slices from 13:29 on 14 January 2008 until 02:09 on 15 January 2008. c) Slices from 02:49 until 16:09 on 15 January 2008 d) Slices from 16:49 on 01/15 /08 until 05:29 on 16 January 2008 e) Slices from 06:09 until 17:29 on 16 January 2008 f) Slices from 18:09 until 20:49 on 16 January 2008.

**Figure 10** Time series of the median of vertical pixels 260:280, at radial pixel 505, 14 – 16 January 2008. We identify a nine-hour (red), four-hour (green), and 2.5 hr (cyan) periodicity with dashes. Letters identify individual plasma enhancements, as identified in Figures 8 and 9.

**Figure 11** Spectral estimate (thick black) of time series shown in Figure 10. Ticks identify significant spectral enhancements in the same colors as Figure 10. The spectral background (thin black) and narrow-band significance thresholds (grey, 95% and 90% confidence thresholds) are shown.

**Figure 12** Illustration of periodic density structures in the ecliptic, and the SECCHI HI1A lines of sight through the structures as a function of distance from the Sun. The solar wind flows from right to left, and periodic enhancements are labeled with letters.

Figure 1

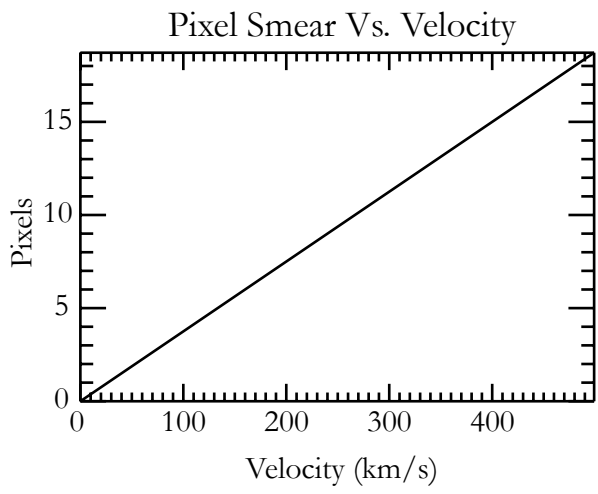

Figure 2

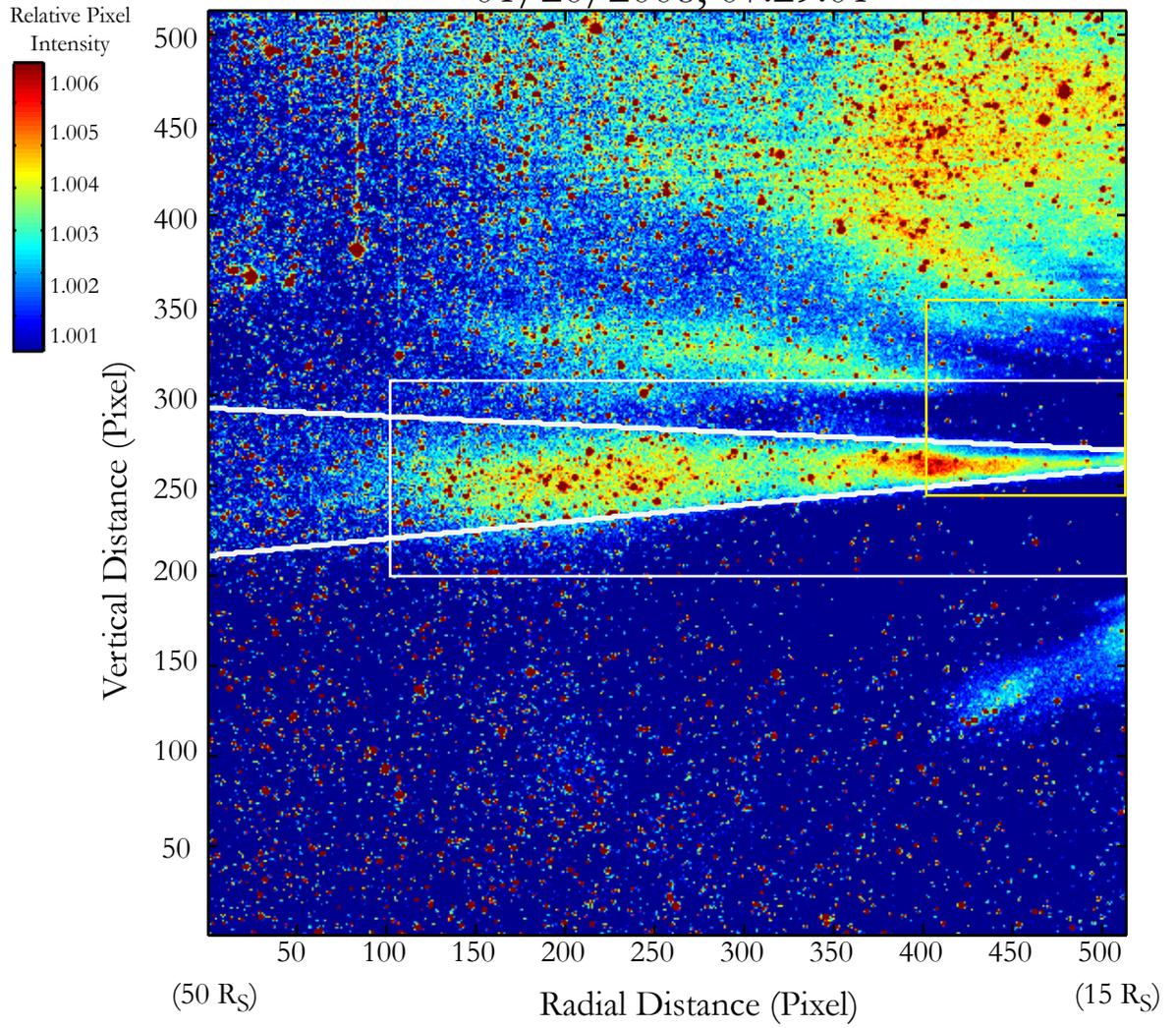

Figure 3

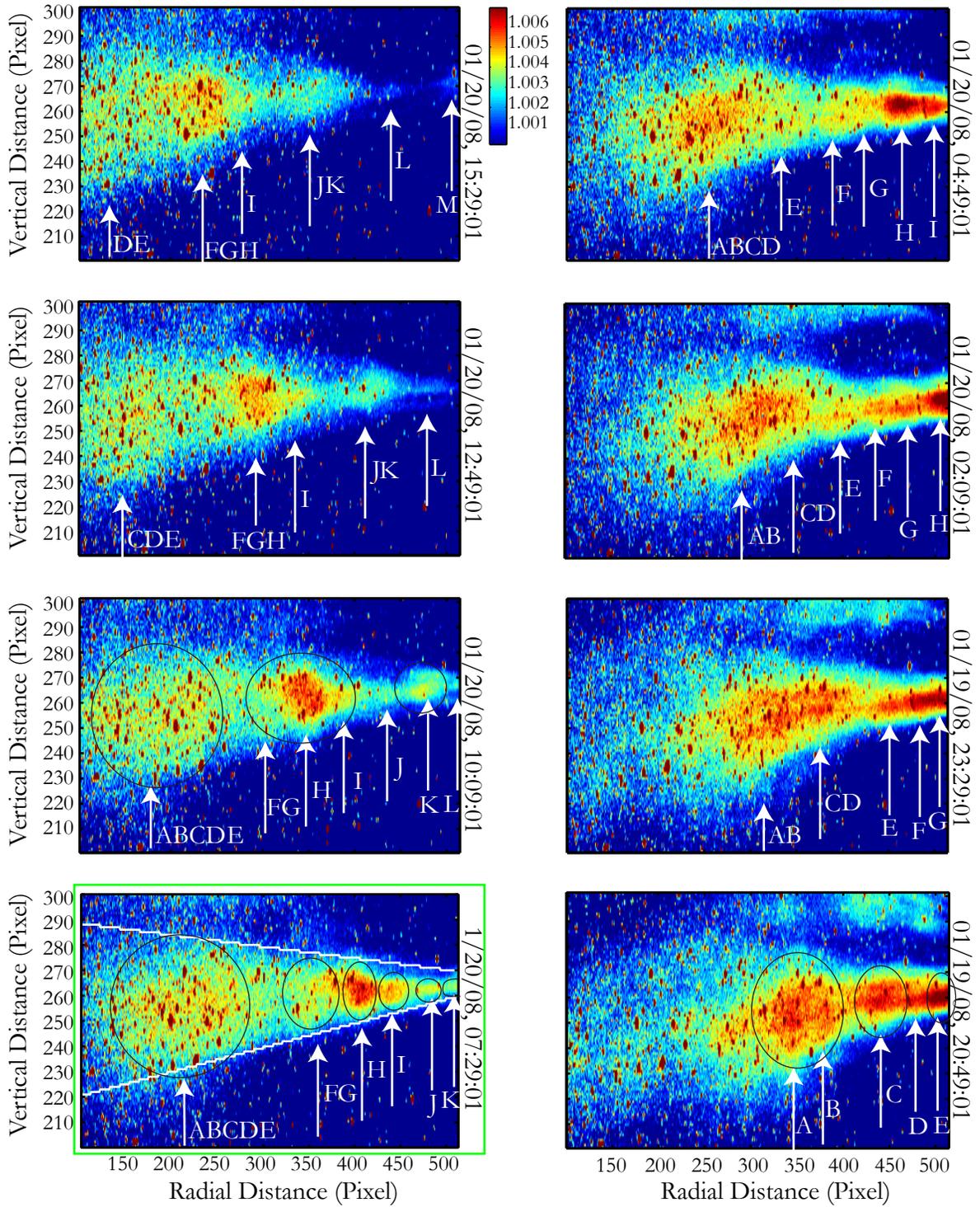

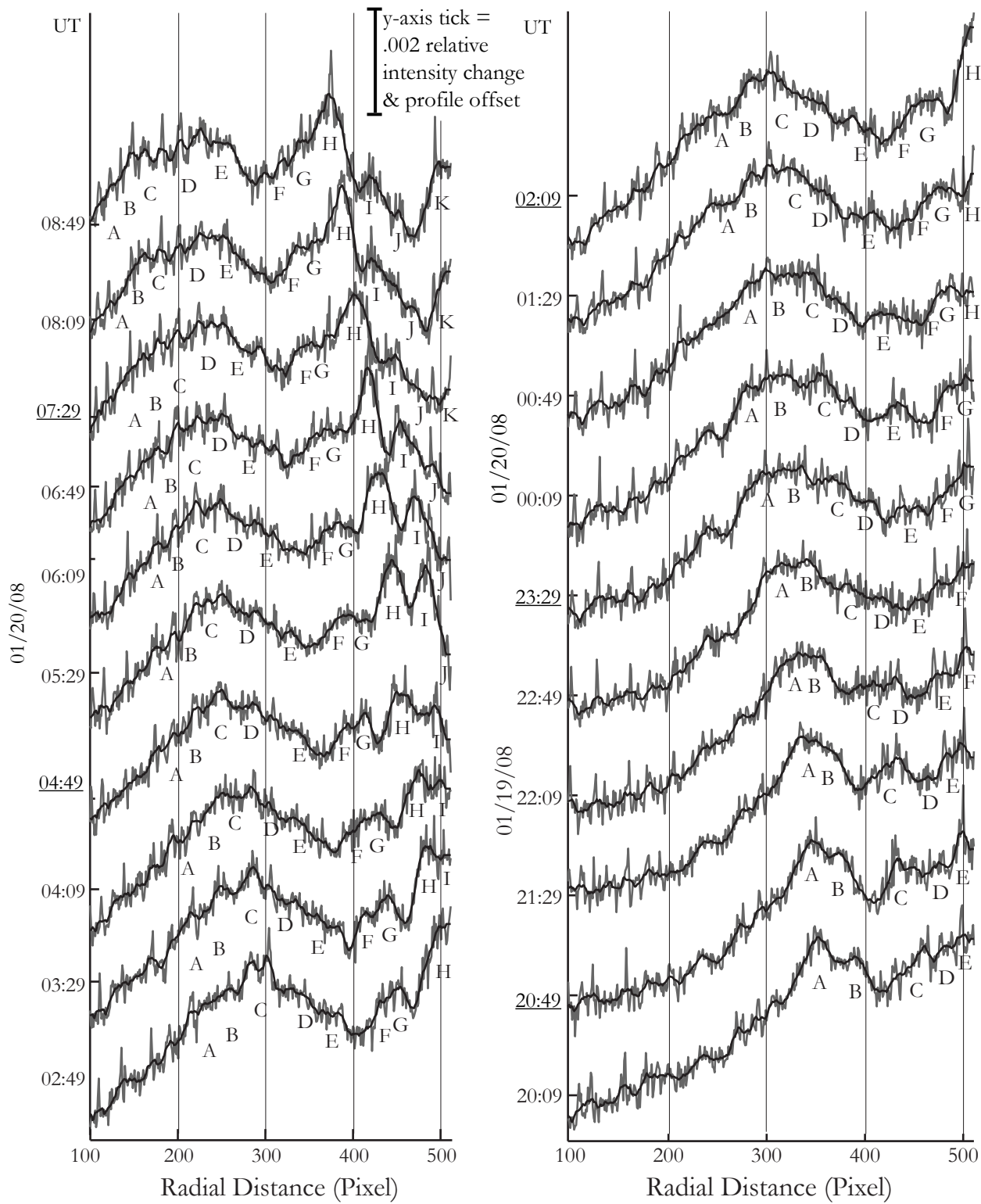

Figure 4a — Event 1, Radial Profiles of Periodic Density Structures

Figure 4b

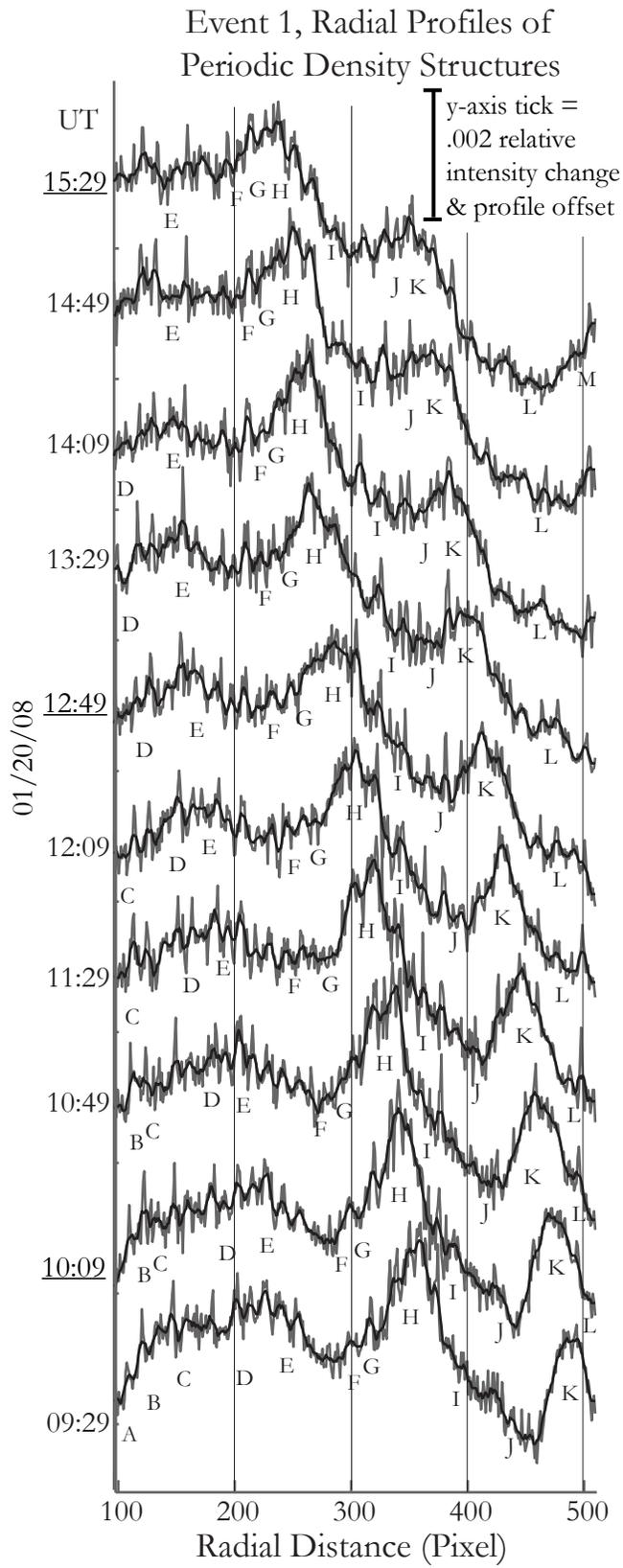

Figure 5

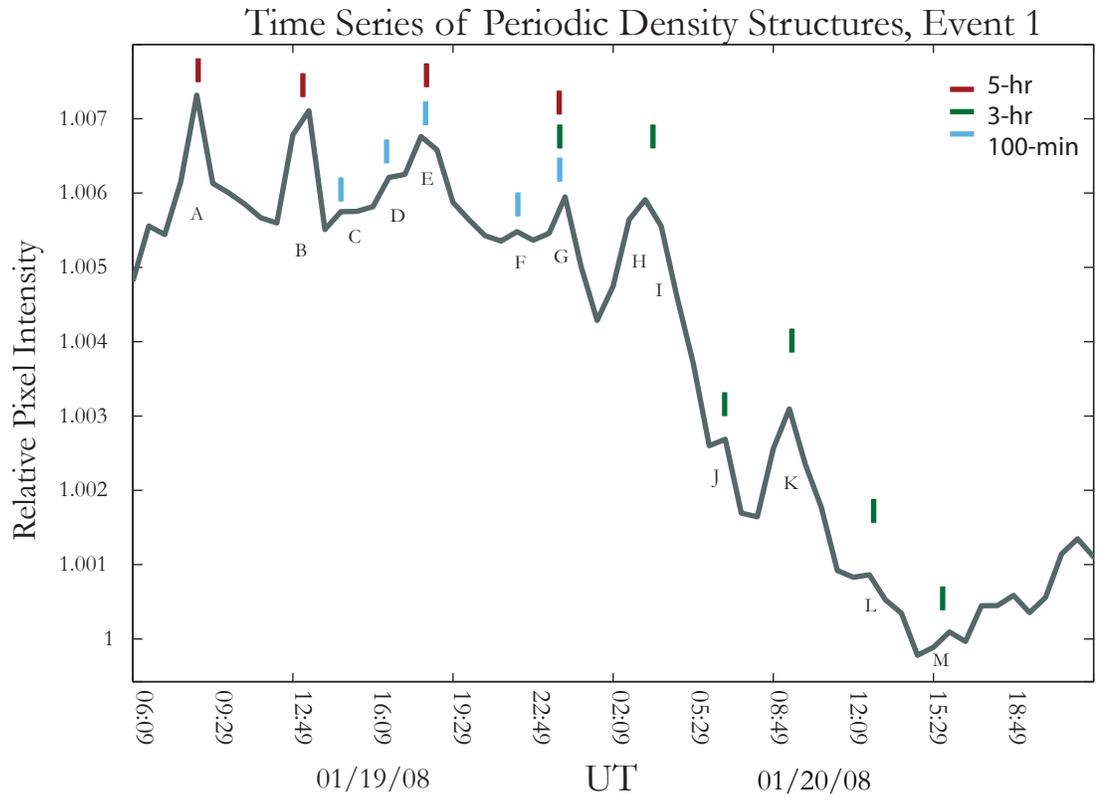

Figure 6

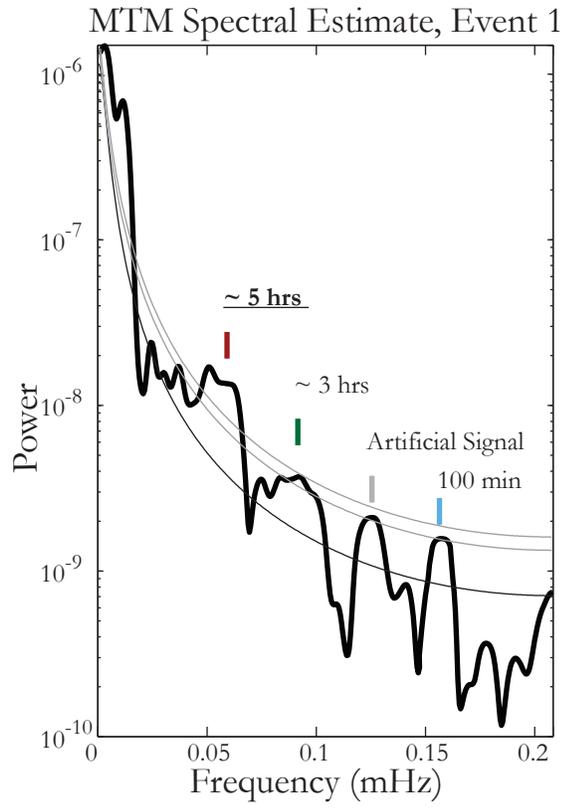

Figure 7

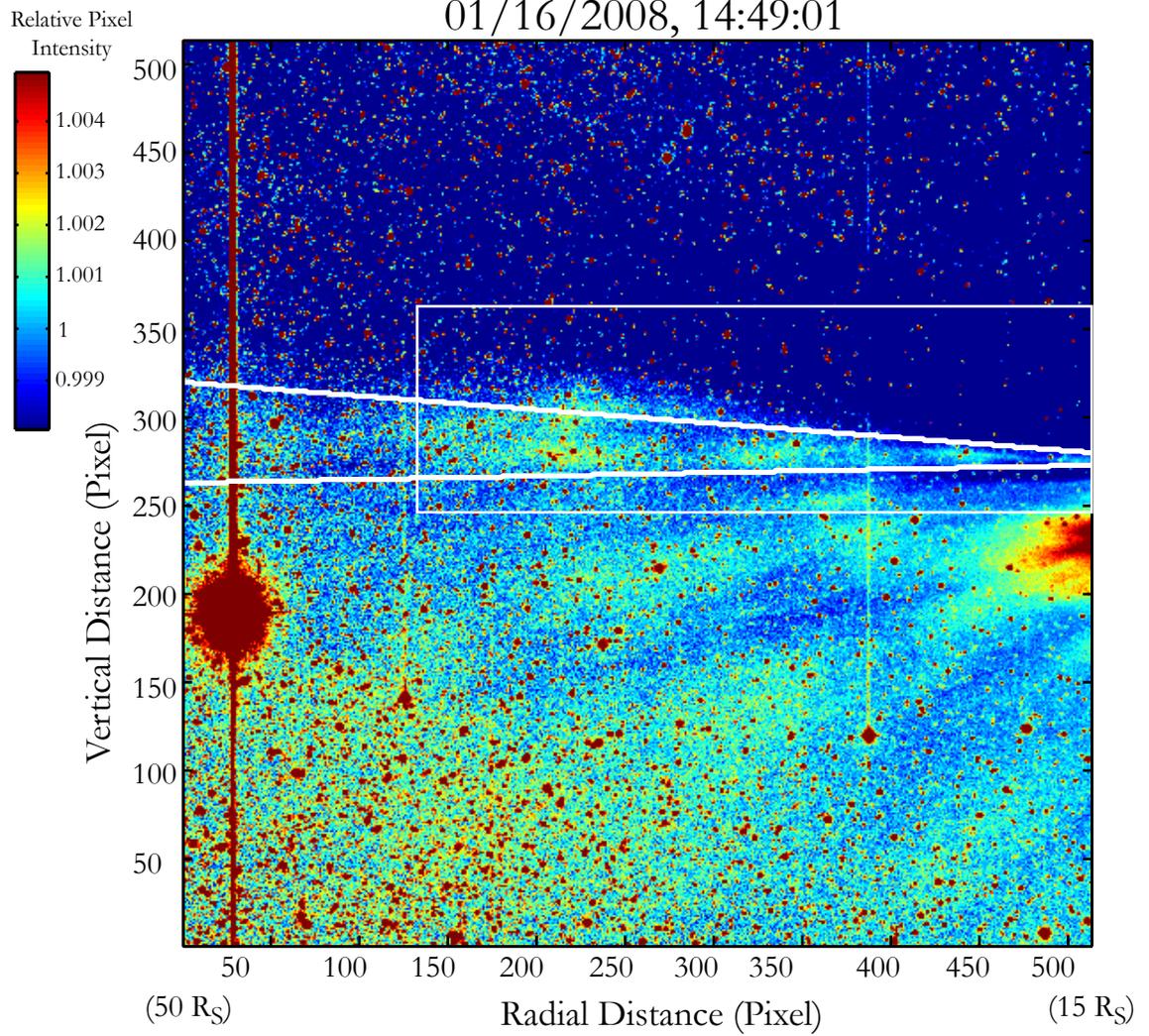

Figure 8a

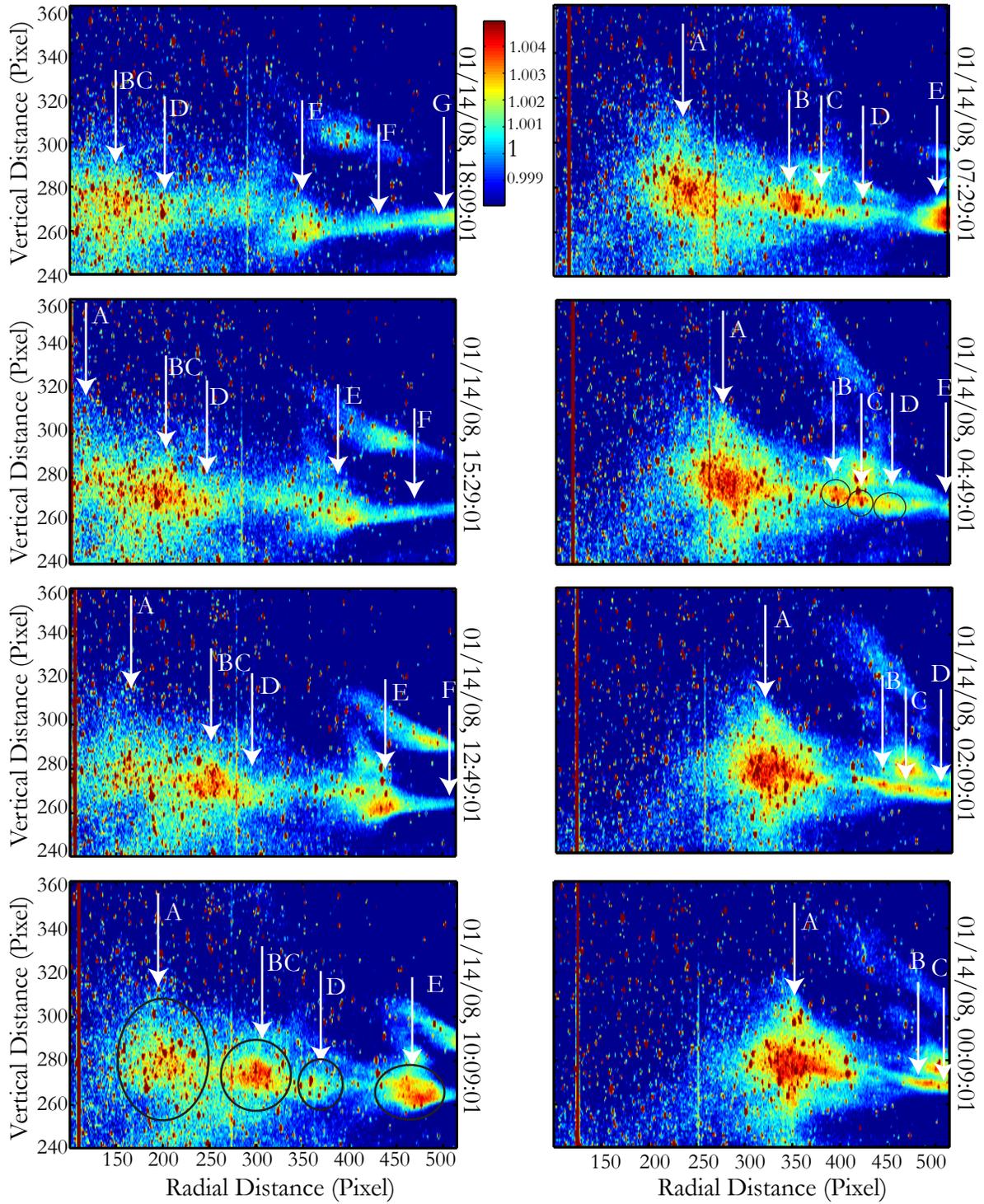

Figure 8b

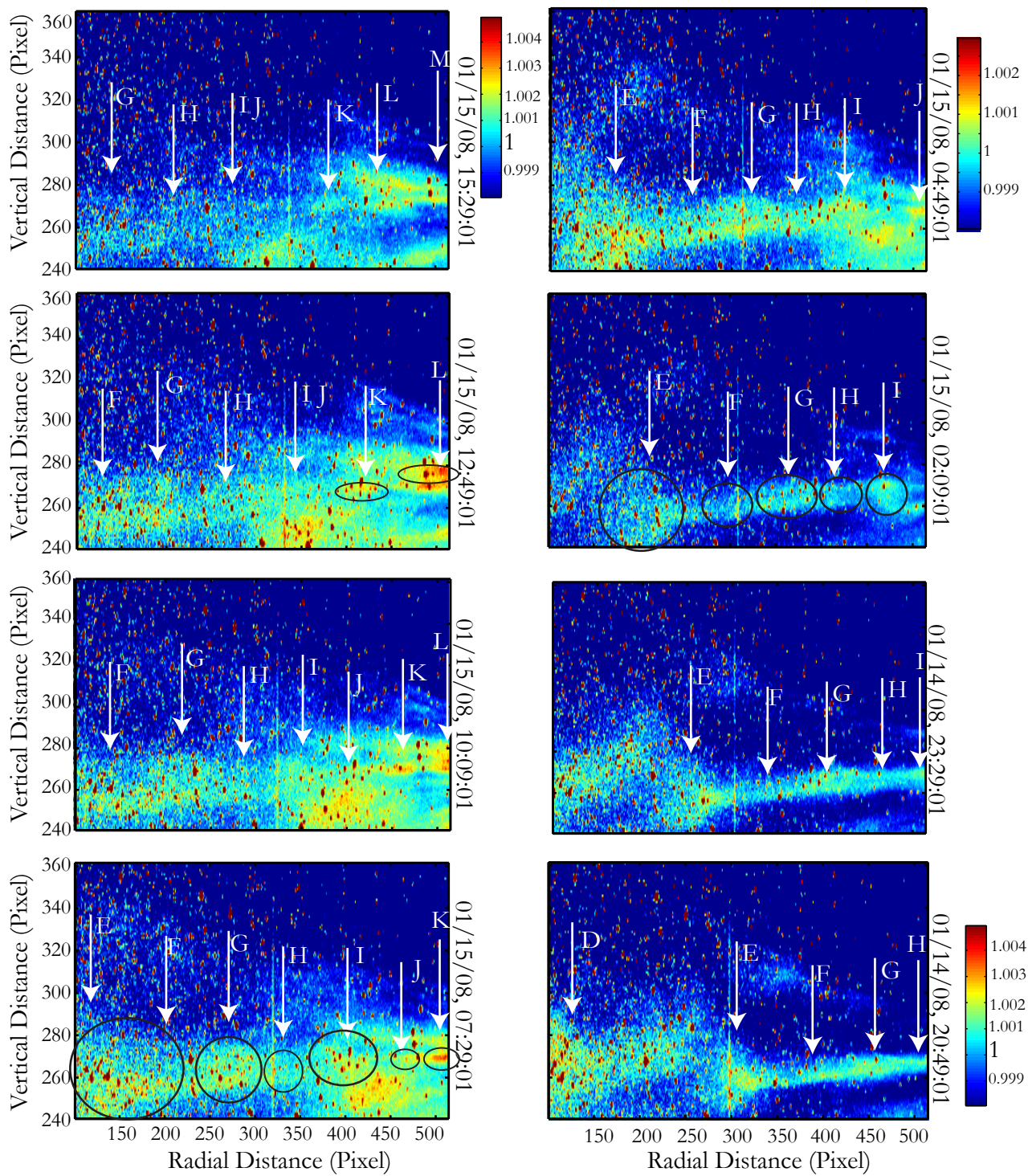

Figure 8c

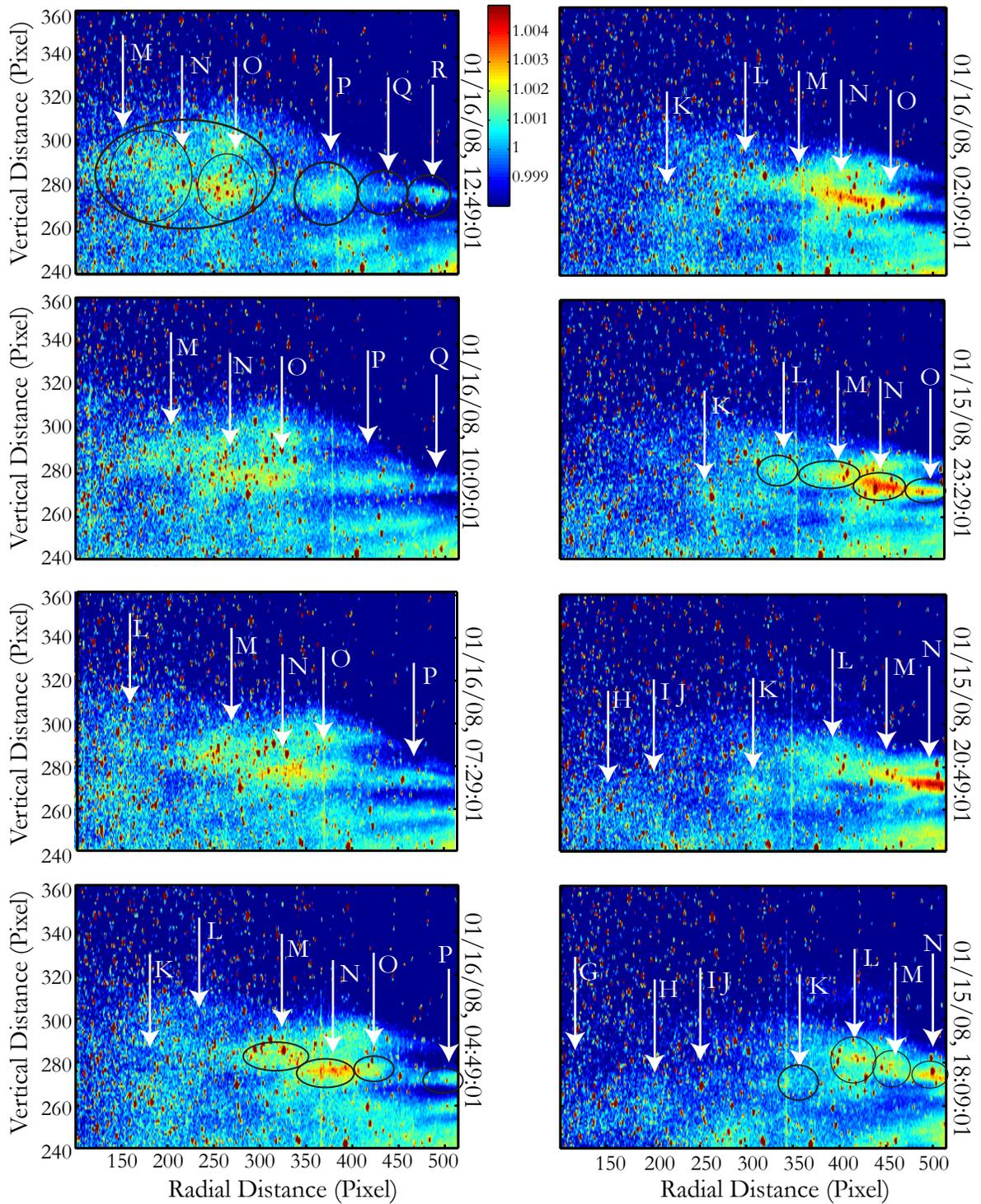

Figure 8d

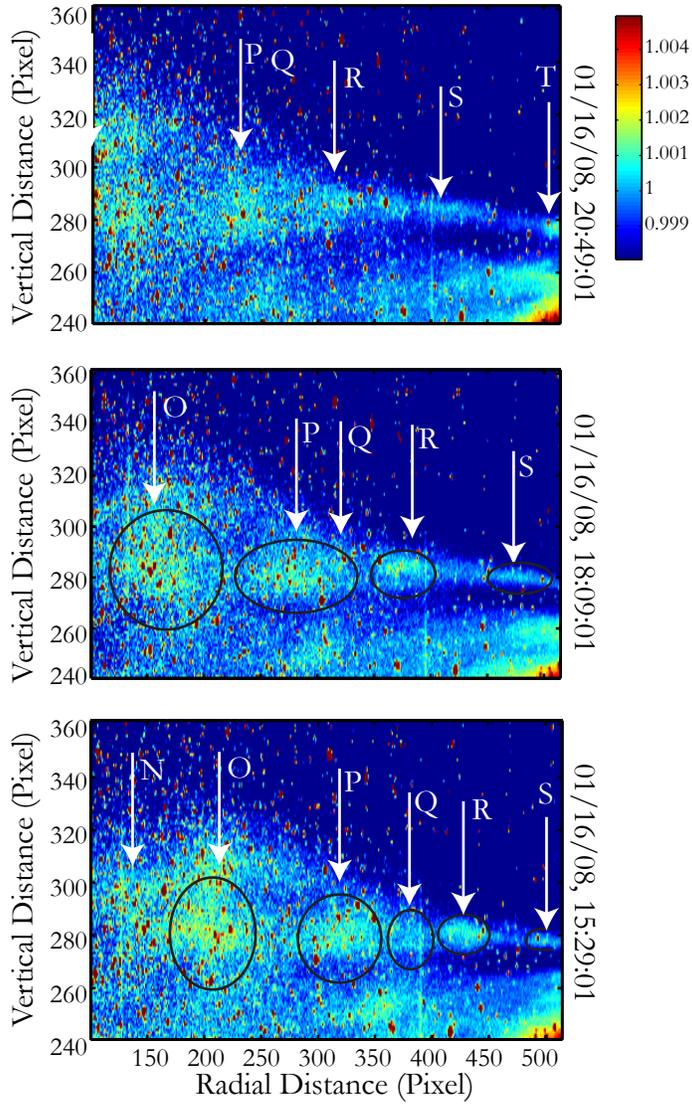

Figure 9a

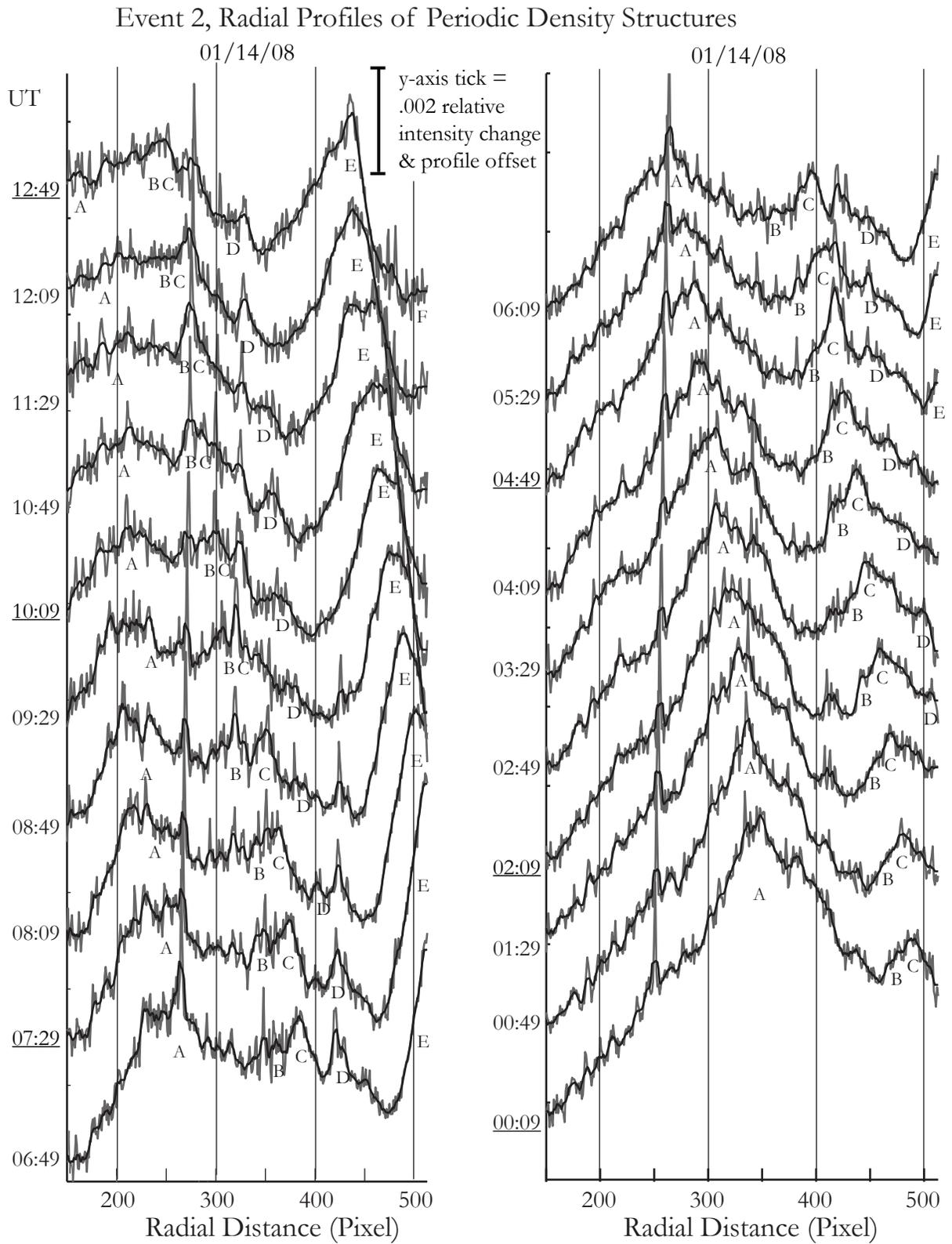

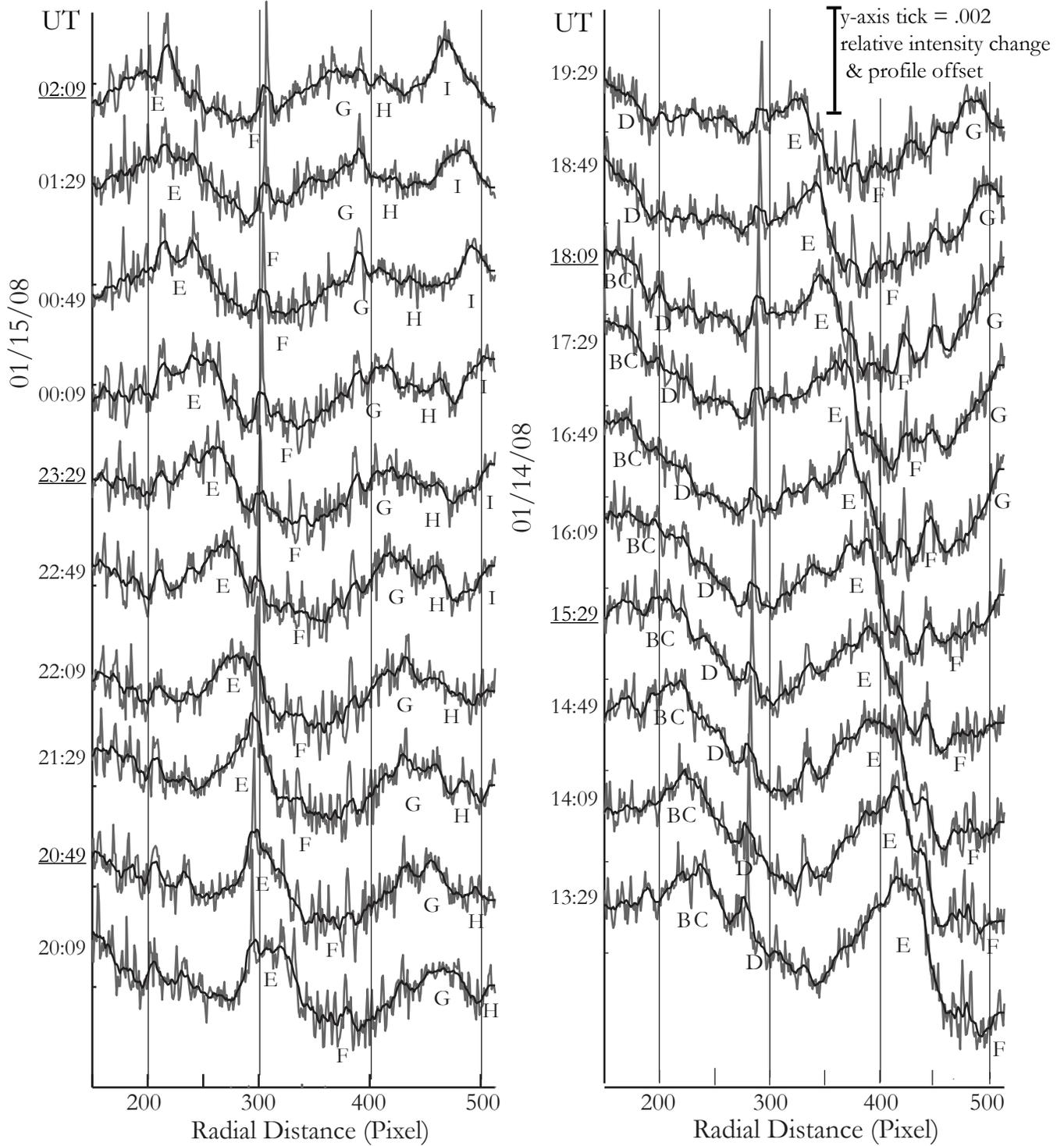

Figure 9c

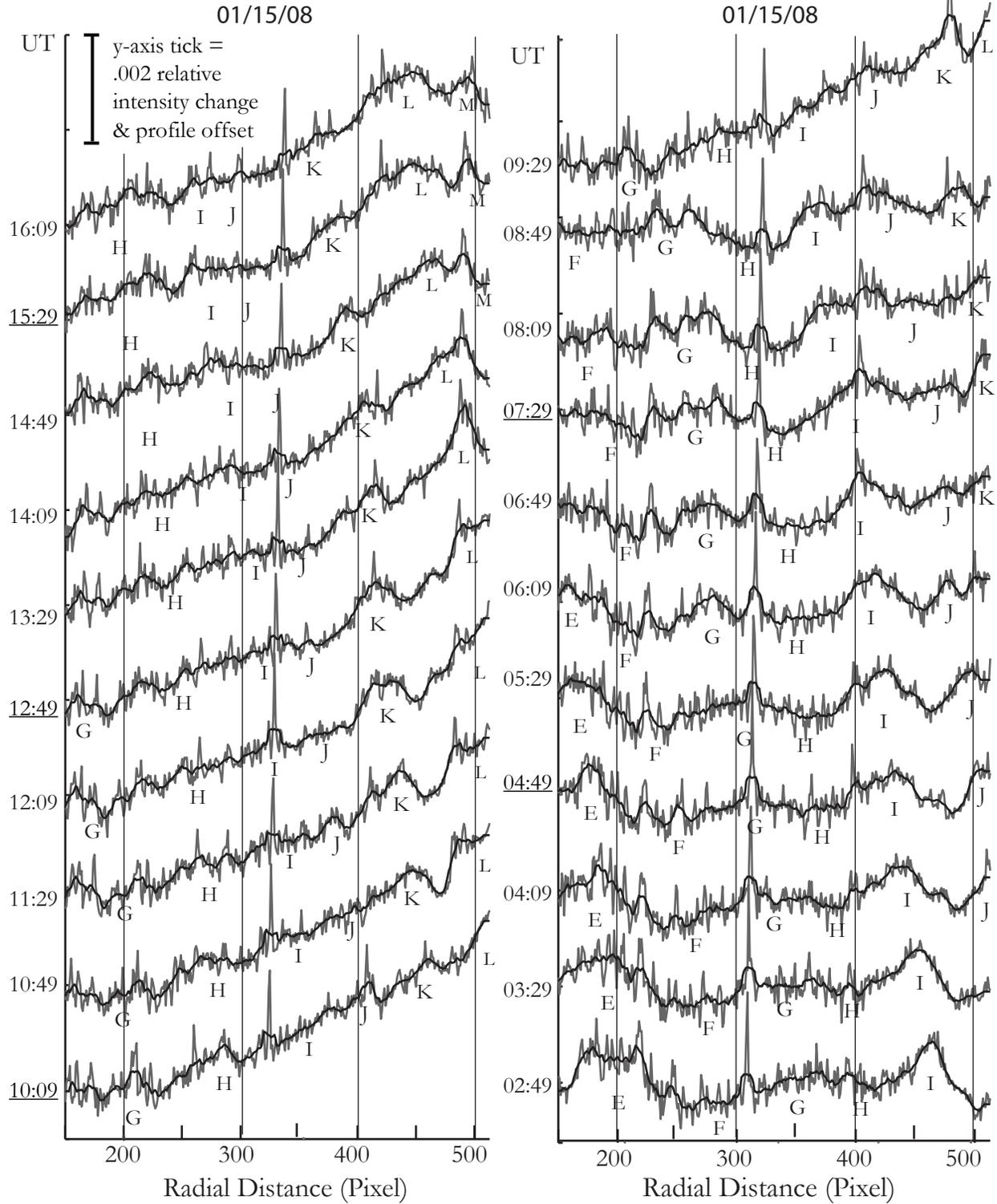

Figure 9d

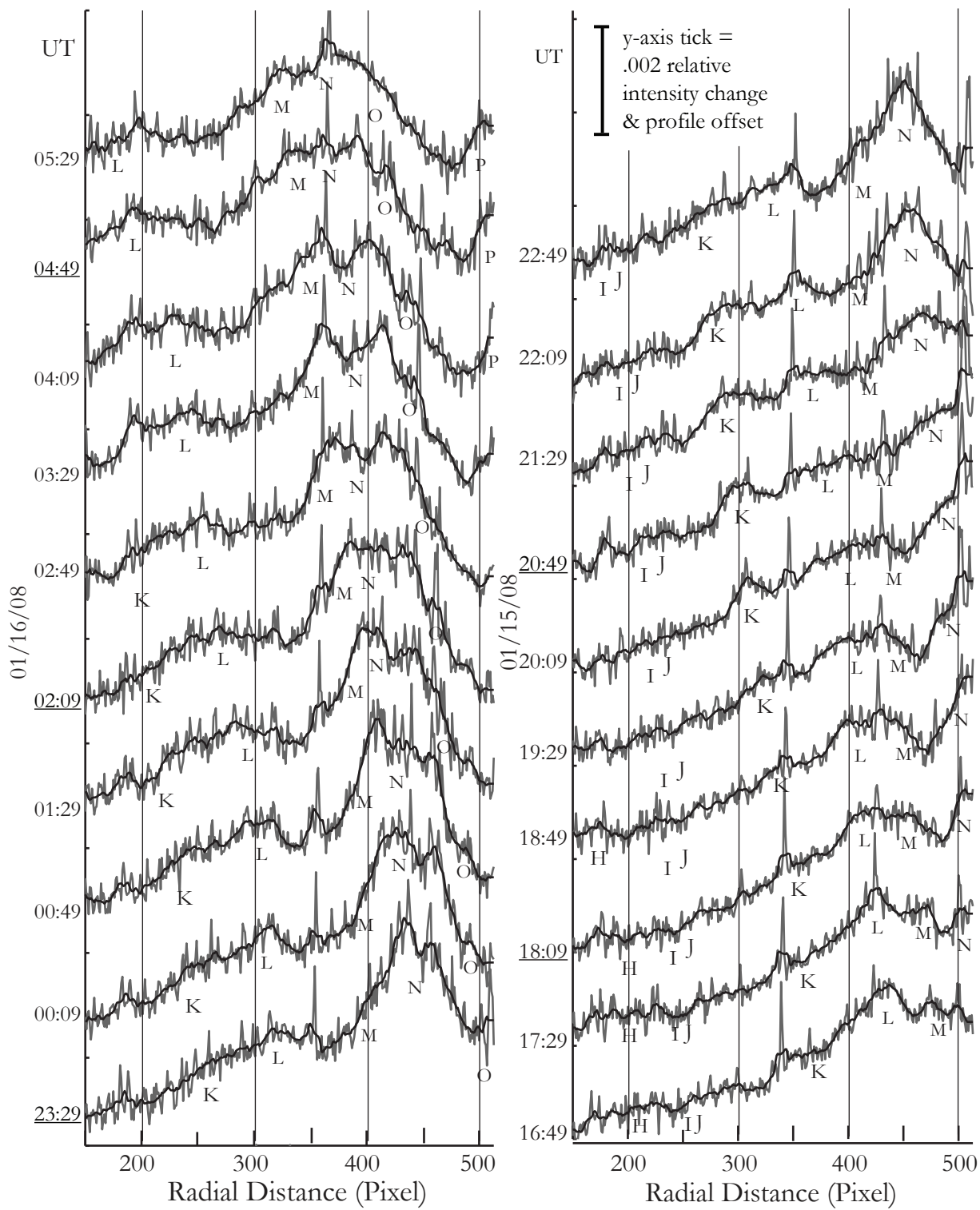

Figure 9e

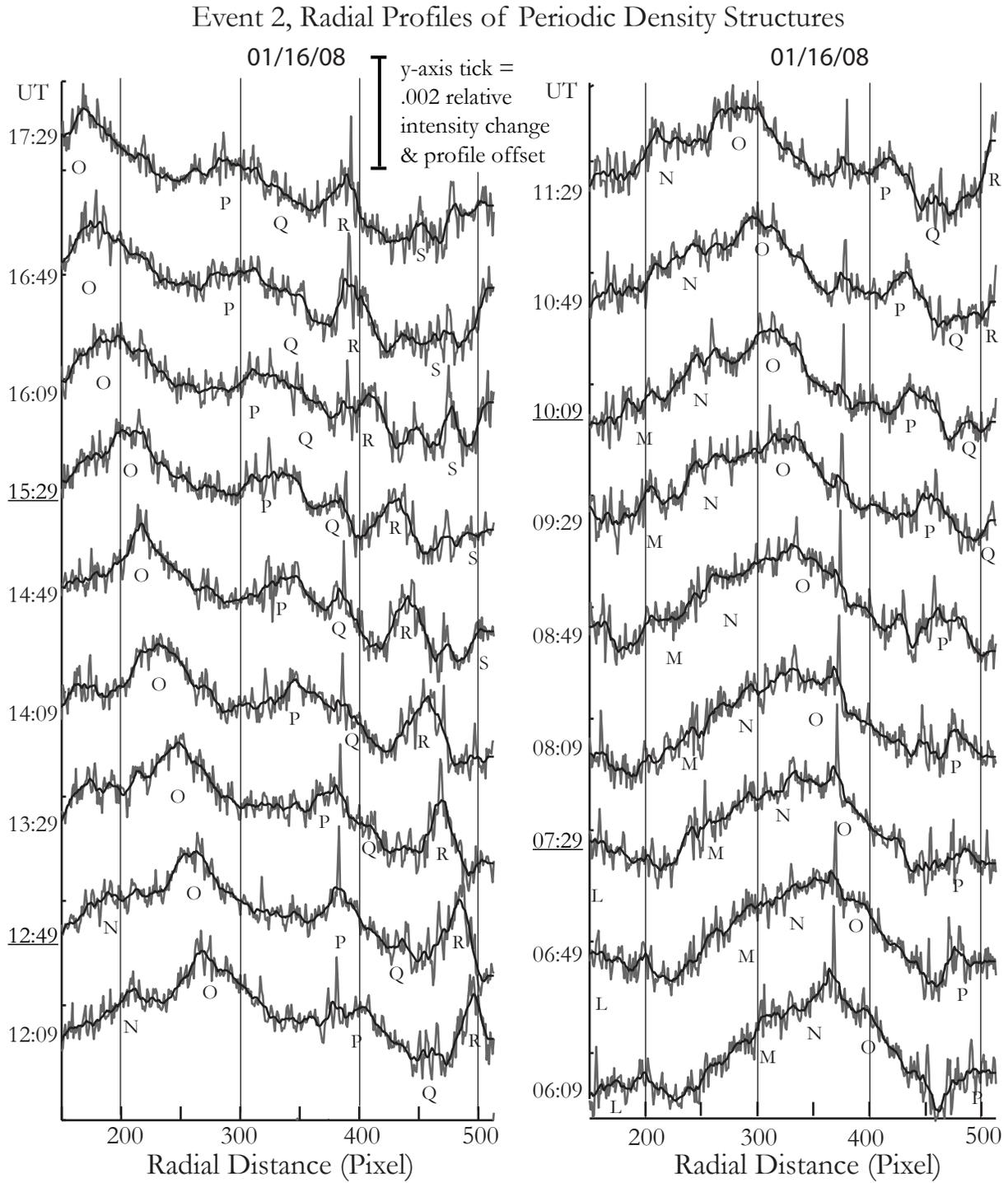

Figure 9f

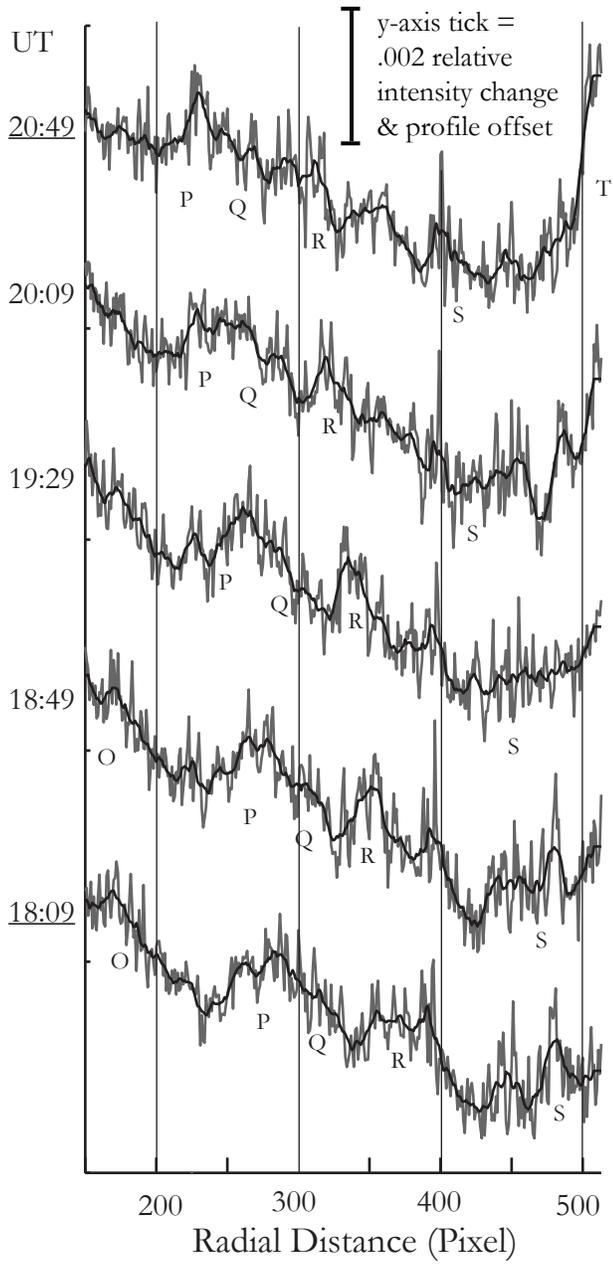

Figure 10

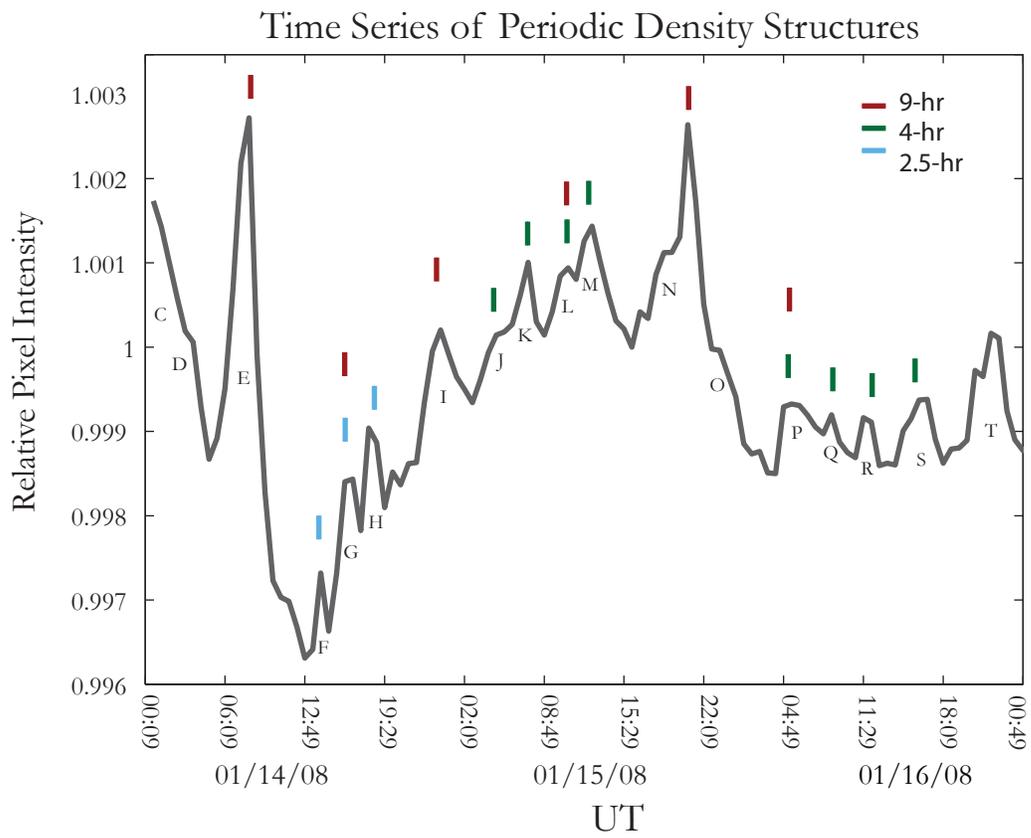

Figure 11

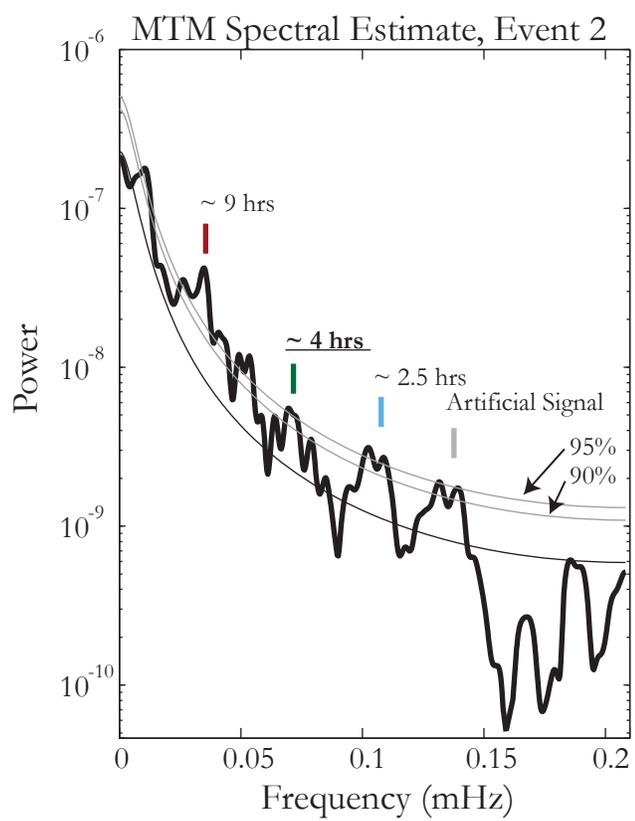

Figure 12

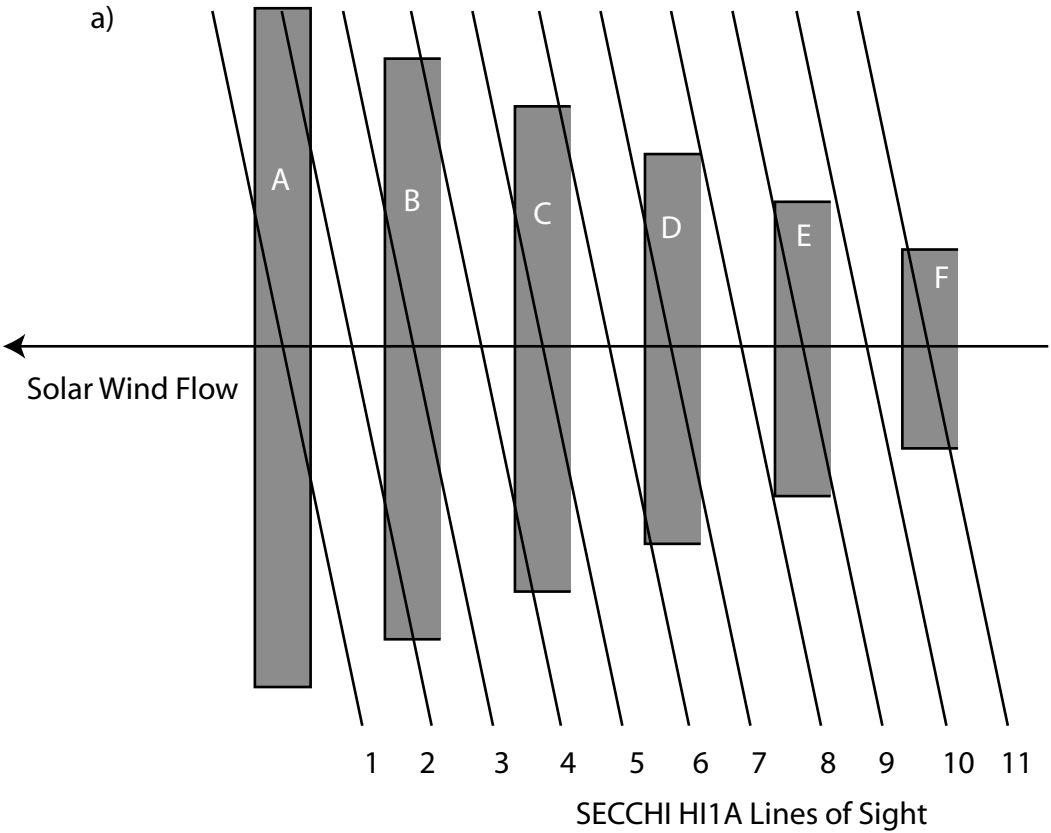

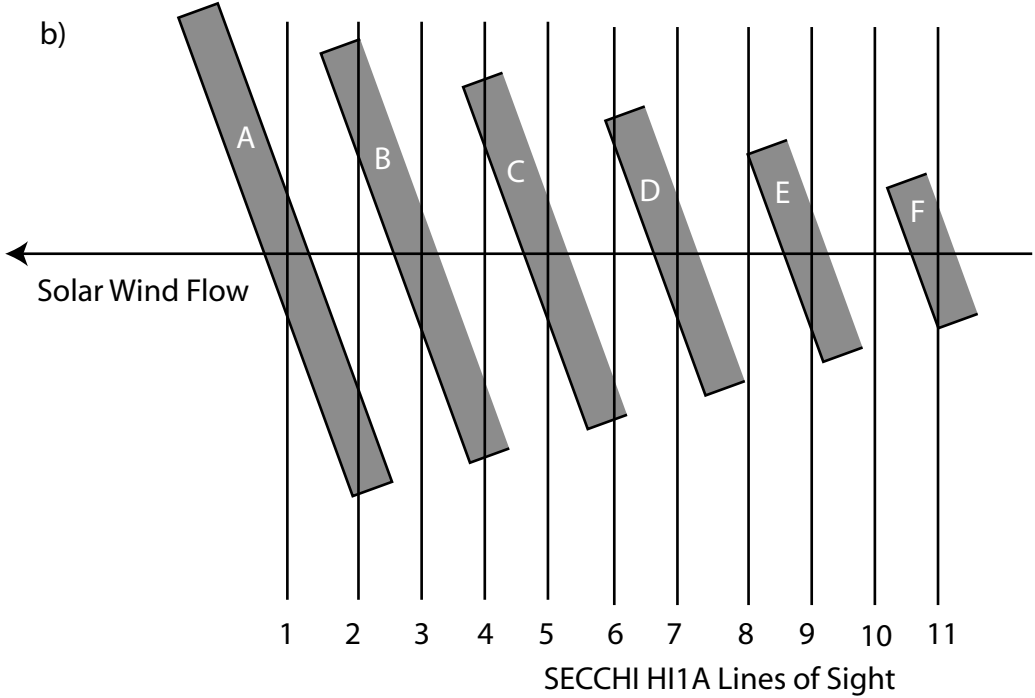